\documentclass[onecolumn,11pt,epsf]{revtex4}

\usepackage{amssymb}
\usepackage{epsfig} 
\usepackage{graphicx}
\DeclareMathAlphabet   {\mathsc}{OT1}{cmr}{m}{sc}  
\def\[{\left [} 
\def\]{\right ]} 
\def\({\left (} 
\def\){\right )}

\newcommand{\lbr}{\left\{} 
\newcommand{\rbr}{\right\}} 
\newcommand{\oline}[1]{\overline{#1}}

\newcommand{\wtd}[1]{\widetilde{#1}} 
\newcommand{\UV}       {\mathsc{uv}} 
\newcommand{\GS}       {\mathsc{gs}} 
\newcommand{\PV}       {\mathsc{pv}}

\newcommand{\hc}       {\mathrm{\; h.c. \;}}

\newcommand{\gappeq}{\mathrel{\rlap {\raise.5ex\hbox{$>$}} 
{\lower.5ex\hbox{$\sim$}}}} 
\newcommand{\lappeq}{\mathrel{\rlap{\raise.5ex\hbox{$<$}} 
{\lower.5ex\hbox{$\sim$}}}}

\begin{document}

\title{Annihilation Radiation of Dark Matter in Heterotic Orbifold Models}

\author{Gianfranco Bertone $^1$, Pierre Bin\'etruy $^{2,3}$, Yann Mambrini $^4$, Emmanuel Nezri $^{2,3}$}

\address{$^1$ NASA/Fermilab Astrophysics Group, Fermi National Accelerator Laboratory, Box 500 Batavia, IL 60510-0500}
\address{$^2$ Laboratoire de Physique Th\'eorique des Hautes Energies, \\ Universit\'e Paris-Sud, F-91405 Orsay}
\address{$^3$ APC Universit\'e Paris 7, Coll\`ege de France, F-75231 Paris Cedex 05}
\address{$^4$ Departamento de F\'{\i}sica Te\'orica C-XI and Instituto de F\'{\i}sica Te\'orica C-XVI, Universidad Aut\'onoma de Madrid, Cantoblanco, 28049 Madrid, Spain}
 
\begin{abstract}
Direct and indirect dark matter searches can potentially
lead to the discovery of Supersymmetry. In the specific context of heterotic 
orbifold compactification scenarios, it is even possible to get some clues on the 
SUSY breaking mechanism of the more fundamental underlying theory. 
In this paper, we investigate the prospects for indirect detection of 
neutralino dark matter via gamma-ray  (continuum and line) and 
synchrotron radiation from the Galactic center, in the context of 
AMSB scenarios in an effective heterotic framework, which constitutes 
a consistent SUSY model between high energy theory and low energy 
phenomenology.

\end{abstract}
\begin{flushright}
LPT Orsay--04/37, \\
FERMILAB-Pub-04/078-A, \\
FTUAM 04/12, \\
IFT-UAM/CSIC-04-28
\end{flushright}
%\tableofcontents
%\clearpage
%\tableofcontents

\maketitle
\section{Introduction}

Many independent measurements provide convincing evidence
that most of the matter in the Universe is dark, i.e. non--baryonic
and of unknown nature, the most common dark matter (DM) candidate being a 
Weakly--Interacting Massive Particle (WIMP)~\cite{Bertone:2004pz}.
Supersymmetric (SUSY) extensions of the Standard Model (SM) naturally 
predict a massive weakly interacting particle (WIMP) called neutralino 
($ \chi^0_1  \equiv \chi$), which, in most versions of SUSY models,
is the lightest supersymmetric particle (LSP), and is made stable by 
virtue of the conservation of R--parity. 
 
Dark matter searches are often discussed in the very popular mSUGRA, 
{\it  a.k.a} CMSSM (constrained MSSM), model ~\cite{mSUGRA} where one assumes unification of 
 the soft SUSY breaking parameters at the GUT scale. This assumption is essentially a
reduction of the parameter space and has no real theoretical motivation. 
More refined SUSY models have been studied in the context of DM searches, 
relaxing the hypothesis of universality
\cite{EllisHiggs,Bottino1,Arnowitt1,Mynonuniv,BirkedalnonU,Nath2,
Profumo:2003em,Ullio:qe,Cesarini:2003nr,Hooper:2003ka, Bottino:2004qi,Cerdeno:2004zj,Chattopadhyay:2004dt,Profumo:2004ty}. 

Here we focus on effective string inspired models, which represent a 
possible consistent framework linking fundamental high energy theory with low 
energy physics. Recently, the full one loop soft supersymmetry breaking terms, 
in a large class of superstring effective theories, have been calculated~\cite{BiGaNe01}, 
based on orbifold compactifications of the weakly--coupled heterotic string 
(including the so--called anomaly mediated contributions). The parameter space 
in this class of models has already been severely constrained taking into 
account accelerator and relic density constraints \cite{Bin1}, direct or
indirect detection of DM from the Sun \cite{Bin2}, or benchmark 
models at the Tevatron \cite{NelsonTeva}. 
In this model, supersymmetry breaking is transmitted by 
the auxiliary ('$F$') fields of the compactification moduli $T^{\alpha}$, 
whose expectation values determine 
the size of the compact manifold, or by the dilaton auxiliary field $S$, whose 
vacuum expectation value determines the magnitude of the coupling constant 
$g_{\mathrm{STR}}$. The spectrum and couplings of SUSY particles depend 
crucially on the contributions of each hidden fields to the breaking
mechanism.

There exist many ways of probing the existence of neutralino dark matter and
supersymmetry.
Of course, accelerator searches are among the most promising ones.
Recent results give a mass limit of 45 GeV for the lightest neutralino, in the 
framework of mSUGRA \cite{LEPWG}. Furthermore, interesting constraints 
have been derived from many experiments on direct and indirect detection of 
dark matter, whose sensitivities are expected to increase by orders of
magnitude in the near future (see Ref.~\cite{Bertone:2004pz} and references therein).
Direct detection is made via the measurement of the recoil energy of
the nuclei of the detector after an elastic scattering with a WIMP.
Indirect detection experiments look for the products of
 annihilation of the LSP in the Sun, Galactic halo or external 
galaxies (see Ref.~\cite{Bertone:2004pz} and references therein).

It has been emphasized in \cite{Bin2} that one--ton detectors are needed for direct searches to test
all the parameter space in scenarios dominated by the dilaton $F-$breaking 
terms, and that moduli dominated scenarios are generically not detectable by dark 
matter searches (direct or indirect detection with neutrinos coming from the Sun) if we combine 
favored relic density and accelerator constraints. 
This means that such DM searches could be a way to distinguish the origin(s) of the SUSY breaking
mechanism, and to restrict or even determine some fundamental 
parameters of the model (of stringy nature or not).

The purpose of the present paper is to investigate DM indirect detection with
high  energy gamma rays and synchrotron
radiation from neutralino annihilation at the Galactic center, in the class of
weakly coupled heterotic 
 string models discussed above \cite{BiGaNe01,Bin1,Bin2,NelsonTeva}.

The paper is organized as follows: after a brief survey on the motivations, 
phenomenology and construction of the effective string models studied here, we analyze
in section 2 and 3 the influence of the different coefficients parameterizing 
the SUSY breaking, on the gamma-ray and synchrotron emission from the Galactic center.
In section 4, we analyze the prospects of observation of gamma-ray fluxes 
with present and  foreseen experiments and discuss synchrotron radiation 
in section 5. We present our conclusions in section 6.

\section{Theoretical framework}

\subsection{Structure of heterotic orbifolds models at one loop}

The task of string phenomenology is  
 to make contact between the high energy string theory, and the low energy 
 world. For this purpose, we need to build a superstring theory in four dimensions, 
 able to reproduce the Standard Model gauge group,  three generations of squarks,
 and a coherent mechanism of SUSY breaking. We set our analysis
 in the framework of orbifold compactifications of the heterotic string, within
 the context of supergravity effective theory, we focus on  
 models where the action is dominated by 1--loop contributions to soft 
 breaking terms. The key property of the models is the non--universality of soft 
 terms, consequences of the beta--function appearing in the superconformal 
 anomalies. This  non--universality gives a peculiar phenomenology in 
 the gaugino and the scalar sector,
 modifying considerably the predictions of mSUGRA models. In fact, these
 string--motivated models show a new behavior, that interpolates between the
 phenomenology of unified supergravity models (mSUGRA) and models dominated by
 the superconformal anomalies (AMSB). The constraints arising from accelerator
 searches, and some dark matter aspects (direct and neutrino indirect
 detections) have been already studied in \cite{Bin1,Bin2}. We extend 
here the analysis to gamma-ray and synchrotron emission from the Galactic center.

We provide a phenomenological study within the context of orbifold
compactifications of the weakly--coupled heterotic string, where we distinguish
two regimes:
\begin{itemize}
\item In the first one, the SUSY breaking is transmitted by the compactification
moduli $T^{\alpha}$, whose vacuum expectation values determine the size of the
compact manifold. Generic (0,2) orbifold models contain three $T_{\alpha}$ 
moduli fields. We considered a situation in which only an "overall modulus $T$"
field contributes to the SUSY--breaking. The use of an overall modulus $T$ is equivalent
to the assumption that the three $T_{\alpha}$ fields of generic orbifold
models give similar contributions to SUSY--breaking. This is expected in the
absence of some dynamical effect that would strongly discriminate
the three moduli.
\item In the second case, it is the dilaton field $S$ present in any
four--dimensional string (whose vacuum
expectation value determine the magnitude of the unified coupling constant
$g_{\mathrm{STR}}$ at the string scale), that transmits, via its auxiliary
component, the SUSY breaking. We work in the context of models in
which string nonperturbative corrections to the K\"ahler potential act to
stabilize the dilaton in the presence of gaugino condensation 
\cite{BiGaWu96,BiGaWu97a}. 
\end{itemize}

The origin of the soft breaking terms are
 completely different in the two scenarios. Some are coming from the
 superconformal anomalies and are non--universal
 (proportional to the beta--function of the Standard Model gauge groups), 
others are generated in the hidden sector (from Green--Schwarz mechanism or
 gaugino condensation) and are thus universal. This mixture between
 universality and non--universality gives the richness of the phenomenology
 in this type of effective string models
 and confirms the interest of non--universal studies in the prospect of
 supersymmetric dark matter detection, non--universality being 
in this case connected with the basic properties of the model.

\subsection{SUSY parameters and constraints}

In all the models considered here, the lightest neutralino is
 the LSP. Its phenomenology (mainly the couplings and the spectrum) is
 determined by its mass matrix, 
\begin{equation}
\arraycolsep=0.01in
{\cal M}_N=\left( \begin{array}{cccc}
M_1 & 0 & -m_Z\cos \beta \sin \theta_W^{} & m_Z\sin \beta \sin \theta_W^{}
\\
0 & M_2 & m_Z\cos \beta \cos \theta_W^{} & -m_Z\sin \beta \cos \theta_W^{}
\\
-m_Z\cos \beta \sin \theta_W^{} & m_Z\cos \beta \cos \theta_W^{} & 0 & -\mu
\\
m_Z\sin \beta \sin \theta_W^{} & -m_Z\sin \beta \cos \theta_W^{} & -\mu & 0
\end{array} \right)\;.
\label{eq:matchi}
\end{equation}

\noindent
written in the ($\tilde{B},\tilde{W}^3,\tilde{H}^0_d,\tilde{H}^0_u$) basis, where, $\tilde B$, $\tilde {W}^3$, $\tilde{H}^0_d$, $\tilde{H}^0_u$ represents
 respectively the B--ino, W--ino and down, up--Higgsinos fields. $M_1$, $M_2$
and $\mu$ are the bino, wino, and Higgs--Higgsino mass parameters respectively.
Tan$\beta$ is the ratio of the $vev$ of the two Higgs doublet fields.
This matrix can be diagonalized by a single unitary matrix $z$ 
such that we can express the LSP as

\begin{equation}
\chi = z_{\chi 1} \tilde B+  z_{\chi 2}\tilde W  
+ z_{\chi 3}\tilde H_1  +z_{\chi 4} \tilde H_2.
\end{equation}

\noindent 

The parameter $\mu$ is obtained under the requirement that the electroweak
symmetry breaking (EWSB) takes place. This is done by computing the complete one--loop
corrected effective potential. The effective $\mu$--term is 
calculated from the EWSB condition after minimization of the scalar
 potential

\begin{equation}
\mu^{2}=\frac{\(m_{H_d}^{2}+\delta m_{H_d}^{2}\) - 
  \(m_{H_u}^{2}+\delta m_{H_u}^{2}\) \tan^2{\beta}}{\tan^{2}{\beta}-1} 
-\frac{1}{2} M_{Z}^{2}
\label{mu}
\end{equation}

\noindent
 The sign of $\mu$ is not determined and
left as free parameter.

 We clearly see that
the nature of the neutralino (and its couplings) depends crucially on the
 hierarchy between the parameters $M_1$, $M_2$ and $\mu$.

\subsection{The moduli dominated scenario}

In the moduli dominated scenario, the one loop order supersymmetric SUSY
 breaking terms at GUT scale can be written \cite{BiGaNe01,GaNeWu99,GaNe00b}:

\begin{eqnarray}
M_a &=& \frac{g_{a}^{2}\(\mu\)}{2} \lbr 2
 \[ \frac{\delta_{\GS}}{16\pi^{2}} + b_{a}
\]G_2(T,\oline{T}) F^{T} + \frac{2}{3}b_{a}\oline{M} \rbr, \label{modsoftgaugi}\\
A_{ijk}&=& - \frac{1}{3} \gamma_{i}\oline{M} - p \gamma_{i} G_2(T,\oline{T})
F^{T} + {\rm cyclic}(ijk), \\ M_{i}^{2} &=& (1-p)\gamma_i
\frac{|M|^2}{9}. \label{modsoftscal}
\end{eqnarray}

\noindent
where $M_a$ and $M_i$ are the soft masses for
 the gauginos and scalars and $A_i$, the trilinear coupling.
$b_a$ is the beta--function coefficient for the gauge group $G_a$:

\begin{equation}
b_a=\frac{1}{16 \pi^2}
\left(
3 C_a - \sum_i C_a^i
\right).
\end{equation}

\noindent
where $C_a$, $C_a^i$ are the quadratic Casimir operators for the group
$G_a$ in the adjoint representation and in the representation of the
field $i$ respectively.
$F^S$ and $F^T$ are the auxiliary fields for the dilaton and the K\"ahler
modulus, respectively, $\overline{M}$ is the supergravity auxiliary fields
whose vacuum expectation value ($vev$) determines the gravitino mass
$m_{3/2}=-\frac{1}{3}\overline{M}$, and $\delta_{\mathrm GS}$ is
the Green--Schwarz coefficient which is a (negative) integer
between $0$ and $-90$. The function  
 $G_2(T, \overline{T})$ is proportional to the Eisenstein function and
vanishes when $T$ is stabilized at one of its two self--dual points.
From Eq.(\ref{modsoftgaugi}), it follows that when the moduli are stabilized at a self dual point, only
the second term contributes to gaugino masses. This is precisely the
"anomaly mediated" contribution.
The loop contributions have been computed using the Pauli--Villars (PV)
regularization procedure. The PV regular fields mimic the heavy string modes
that regulate the full string amplitude. The phenomenological parameter $p$
which represents the effective modular weight of the PV fields is constrained to be not larger than $1$, though it can be negative in
value. Thus the scalar squared mass for all matter fields is in
general non--zero and positive at one loop (only the Higgs can have a
negative running squared mass). The limiting case of $p=1$, where the
scalar masses are zero at one loop level and for which we recover a sequestered
sector limit, occurs when the regulating PV fields and the mass--generating
PV fields have the same dependance on the K\"ahler moduli. Another reasonable
possibility is that the PV masses are independent of the moduli, in which
case we would have $p=0$.
and $\gamma_i$ is related to the anomalous dimension through 
$\gamma_i^j=\gamma_i \delta_i^j$.
(see \cite{BiGaNe01,Bin1,NelsonTeva} for notations and conventions)

\begin{figure}
\begin{center}
\begin{tabular}{cc}
 \includegraphics[width=0.45\textwidth]{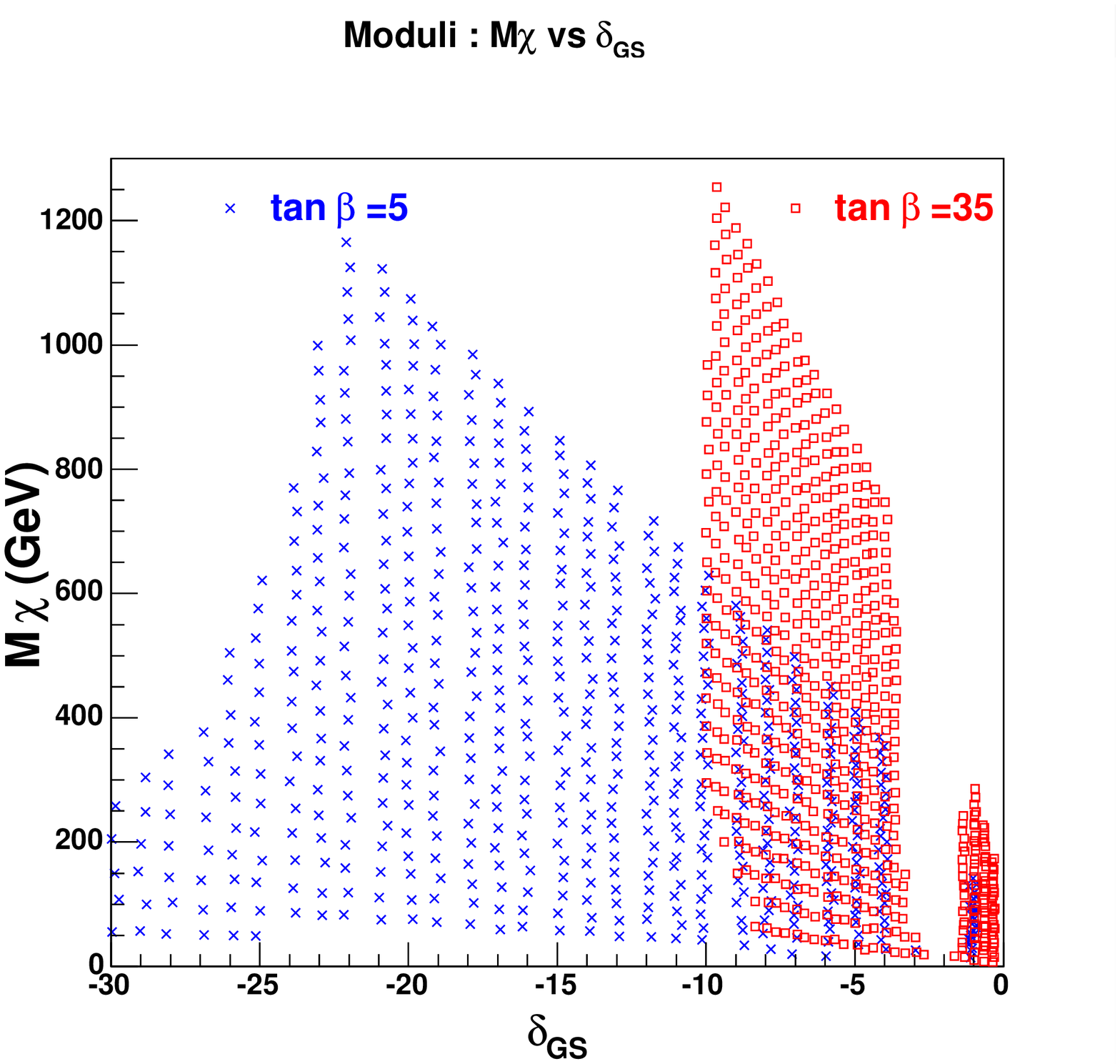}&
\includegraphics[width=0.45\textwidth]{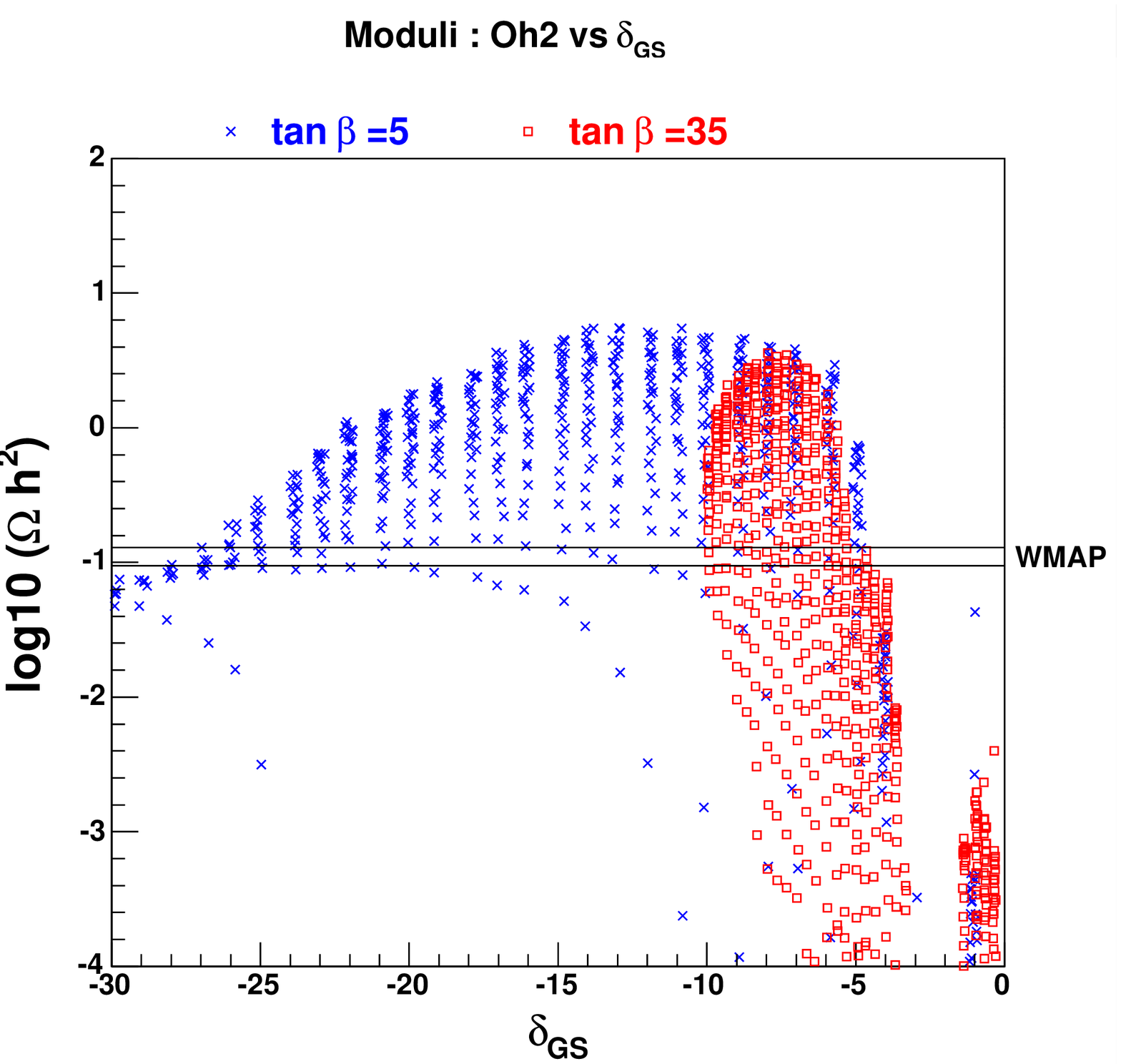}\\
a) & b)
\end{tabular}
\caption{\small a)  Neutralino mass b) Relic density as a function of
  $\delta_{GS}$ for $t=0.25$, $p=0$, $\tan{\beta}=5$ in blue crosses
  (resp. 35 in red boxes) and
  0<$m_{3/2}$<10 (resp. 20) TeV. The hole for small $|\delta_{GS}|$ values
  corresponds to a gluino LSP corridor and the left boundary of the clouds
  corresponds to the stau LSP region (see Figs.\ref{fig:BR-modtb35} and \ref{fig:BR-modtb5})}
\label{fig:mchi-Oh2-dGS}
\end{center}
\end{figure}

We clearly see in these formulae the competition between universal terms and
non--universal ones. The scalar mass terms are all non--universal 
and proportional to their anomalous dimension  and thus loop suppressed.
 The Green-Schwarz mechanism generates universal breaking terms for
the gauginos (proportional to $\delta_{\mathrm GS}$) whereas superconformal
 anomalies introduce non--universal contributions (proportional to $b_a$).
 The nature of the neutralino thus depends mainly on the value 
of the Green--Schwarz counterterm $\delta_{\mathrm GS}$, whereas the mass
 scale  is the gravitino mass $m_{3/2}$.  
We illustrate in Figs.\ref{fig:mchi-Oh2-dGS} the
neutralino mass and its relic density as a function of $\delta_{GS}$, 
vertical directions correspond to different values of $m_{3/2}$.

\subsection{The dilaton dominated scenario}

We turn now to a scenario where the dilaton is the primary source
of supersymmetry breaking in the observable sector. It is well known
that if we use the standard K\"ahler potential derived from the tree
level string theory, it is very difficult to stabilize the dilaton at
 acceptable weak--coupling values. We postulate, in our study, nonperturbative
correction of stringy origin to the dilaton K\"ahler potential. In that case, 
one condensate can stabilize the dilaton at weak coupling, while simultaneously
ensuring vanishing expectation values at the minimum of the potential.
The key feature of such models is the deviation of the dilaton K\"ahler
metric from its tree level value. If we imagine the superpotential for the
 dilaton having the form $W(S) \propto e^{-3S/b_+}$, with
$b_+$ being the largest beta--function coefficient among the condensing
gauge groups of the hidden sector, then we are led to consider the
 phenomenology of models given by the following pattern of soft supersymmetry
 breaking terms \cite{BiGaNe01,GaNeWu99,GaNe00b}:

\begin{eqnarray}
M_{a}&=&\frac{g_{a}^{2}\(\mu\)}{2} \lbr  \frac{2}{3}b_{a}\oline{M}
+\[ 1 - 2 b_{a}' K_s \] F^{S} \rbr \label{dilatsoftgaugi}\\ A_{ijk} &=&
-\frac{K_s}{3}F^S - \frac{1}{3} \gamma_{i}\oline{M} +
\tilde{\gamma}_{i} F^{S} \lbr \ln(\mu_{\PV}^{2}/\mu_R^2)
-p\ln\[(t+\bar{t}) |\eta(t)|^4\] \rbr + (ijk) 
 \\ M_{i}^{2} &=& \frac{|M|^2}{9}
 \[ 1 + \gamma_i
-\(\sum_{a}\gamma_{i}^{a} -2\sum_{jk}\gamma_{i}^{jk}\) \(
\ln(\mu_{\PV}^{2}/\mu_R^2) -p\ln\[(t+\bar{t}) |\eta(t)|^4\] \) \]
\nonumber
 \\
 & &+ \lbr
 \wtd{\gamma}_{}\frac{MF^S}{6}+\hc \rbr , \label{dilatsoftscal}
\end{eqnarray}

\noindent
where $\mu_{\UV}$ is an ultraviolet regularization scale (of the order of 
the string scale $M_{\mathrm{STR}}$) and $\mu_R$ the renormalization scale (taken at the
boundary value of $M_{\mathrm{GUT}}$)\footnote{For simplicity, we assume here
that $M_{\mathrm{STR}} \sim \mu_{\UV} \sim \mu_R \sim M_{\mathrm{GUT}}$
(the corresponding error is logarithmic and appears in a loop factor).}. Moreover

\begin{equation}
F^{S} = \sqrt{3} m_{3/2} (K_{s\bar{s}})^{-1/2}, ~~~~ 
K_{s\bar{s}} = \partial_s \partial_{\overline{s}} K ,
\label{FS}
\end{equation}

\noindent
and

\begin{equation}
(K^{s\bar{s}})^{-1/2} = \sqrt{3}
\frac{\frac{2}{3}b_{+}}{1-\frac{2}{3}b_{+}K_{s}}, ~~~~
K_{s}=-g_{\mathrm STR}^2/2 . 
\label{Ktrue}
\end{equation}

\noindent
(see \cite{BiGaNe01,Bin1,NelsonTeva} for notations and conventions)
to ensure a vanishing vacuum energy in the dilaton--dominated limit.

\begin{figure}
\begin{center}
\begin{tabular}{cc}
 \includegraphics[width=0.45\textwidth]{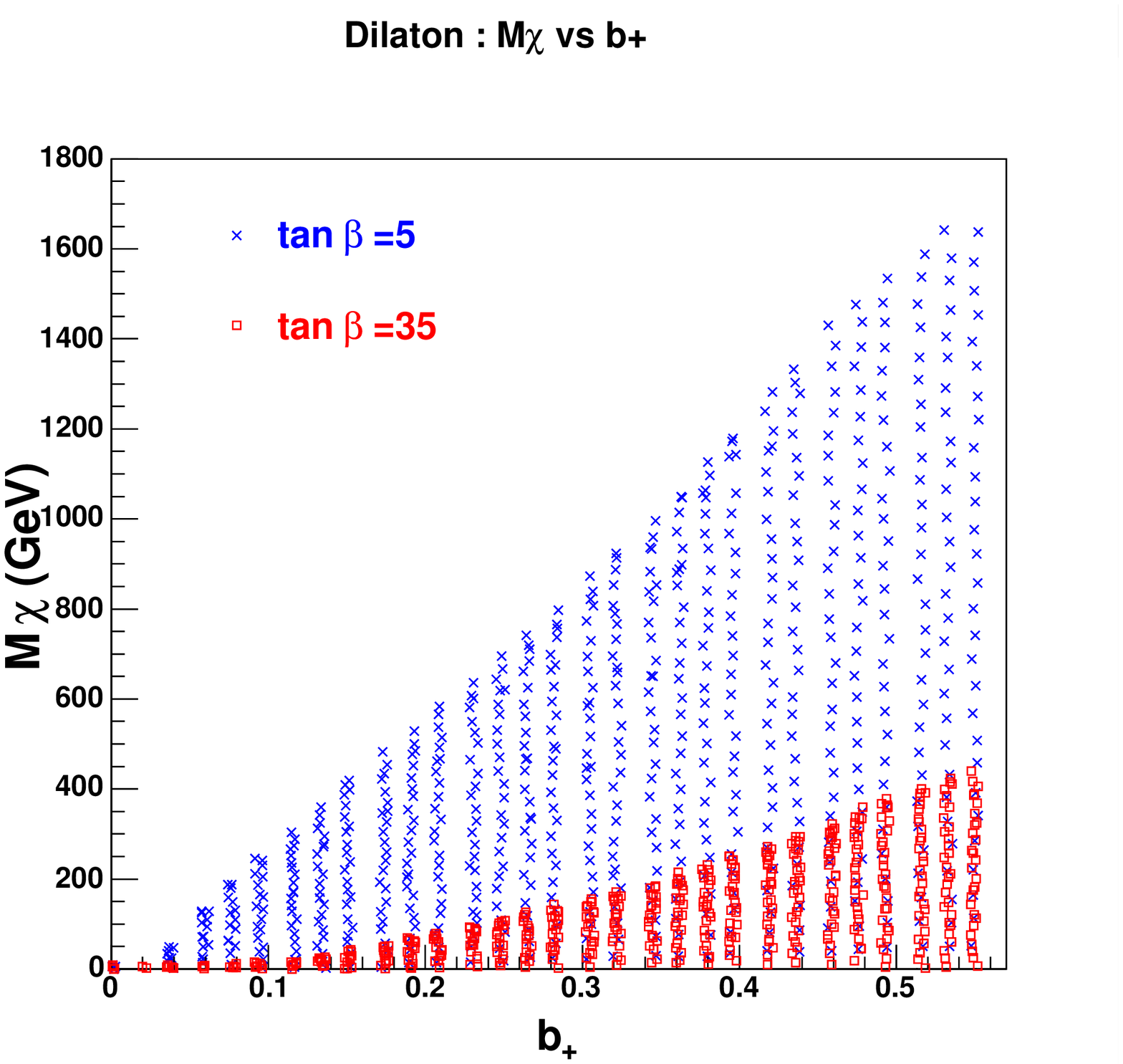}&
\includegraphics[width=0.45\textwidth]{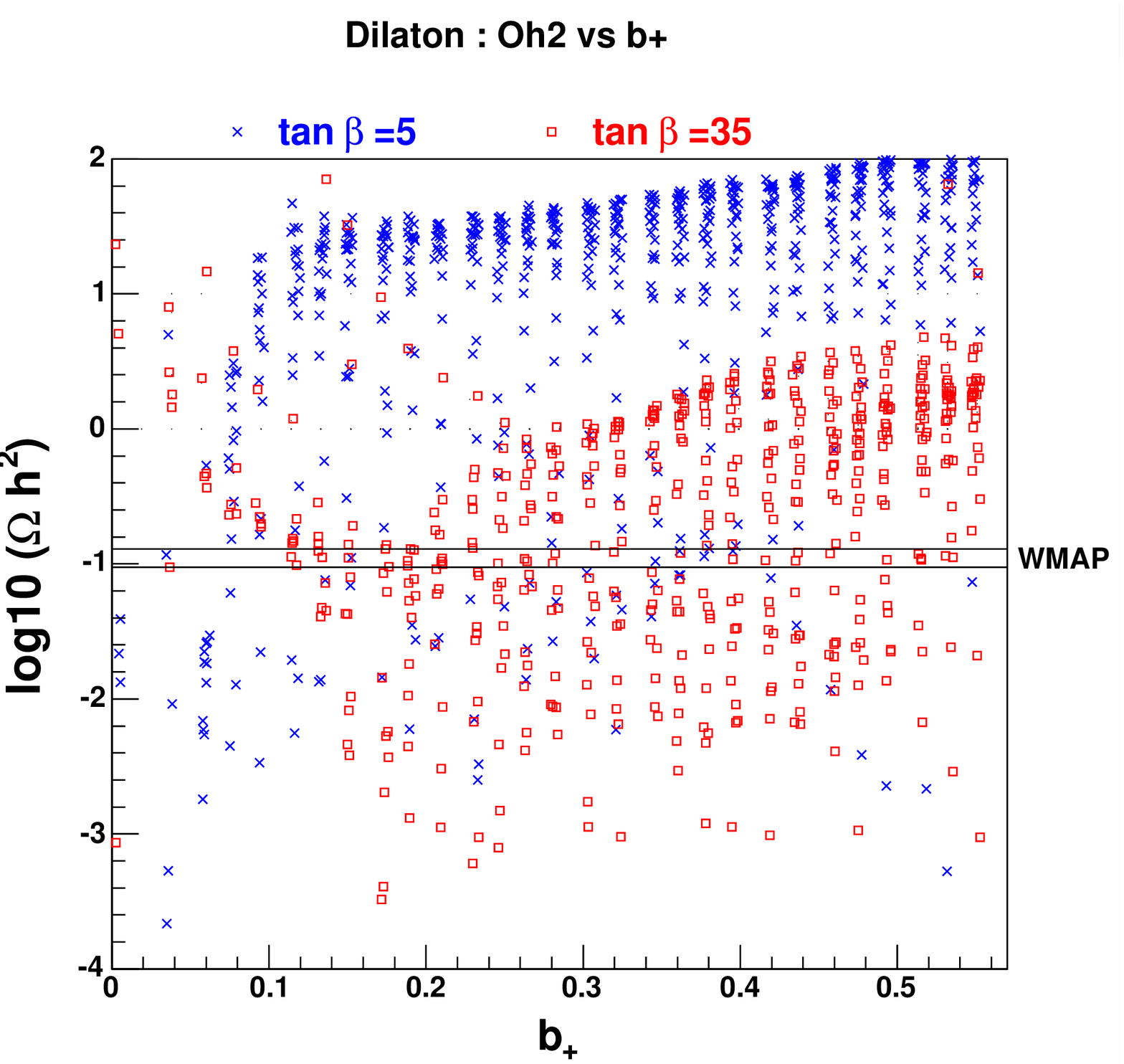}\\
a) & b)
\end{tabular}
\caption{\small a)  Neutralino mass b) Relic density as a function of $b_{+}$
  for $t=0.25$, $p=0$, $\tan{\beta}=5$  in blue crosses (resp. 35 in red boxes) and
  0<$m_{3/2}$<15 (resp. 4) TeV.}
\label{fig:mchi-Oh2-bplus}
\end{center}
\end{figure}

The phenomenology of a dilaton--dominated scenario is completely different
from a moduli--dominated one. 
If we look at formulae (\ref{dilatsoftgaugi}) and (\ref{dilatsoftscal}),
 it is clear that we are in a domain of heavy squarks and sleptons 
(of the order of magnitude of the gravitino mass) 
and relatively light gauginos. 
Indeed, the factor $b_+$, as it contains a loop factor, can suppress
the magnitude of the auxiliary field $F^S$ relative to that of the 
supergravity auxiliary field $M$ through the relation (\ref{FS}).
The resulting gaugino soft breaking terms are less universal for low values 
of $b_+$.  
%On the other hands, for very large values of $b_+$, $F^S$ is roughly
%proportional to $b_+$ and is the dominant contribution to the gauginos
%soft breaking terms. Thus, a large $b_+$ regime is equivalent to 
%a mSUGRA scenario with $M_0 \sim m_{3/2}$ and 
%$M_{1/2} \sim g^2_{\mathrm STR} b_+ m_{3/2} \sim b_+ m_{3/2} / 2$.   
We illustrate neutralino mass and relic density as a function of
 $b_+$ in Figs.\ref{fig:mchi-Oh2-bplus}.

\section{Supersymmetric dark matter phenomenology}

\subsection{Relic density}

The relic density of neutralinos depends on their composition. 
In a large parameter space of the mSUGRA model, the bino--like nature of the 
lightest neutralino $\chi$ implies a rather low rate of annihilation.
% and
%as a direct consequence, an over--abundance of the dark matter density.
 
Different processes lead to interesting neutralino relic
density. Let us begin with a brief survey. For a bino neutralino one needs
sfermion coannihilation or
annihilation into the pseudo-scalar $A$ to have a cosmologically favoured
abundance of neutralino. If the lightest neutralino has a 
dominant wino component the relic density drops 
because of
efficient annihilations into gauge bosons as well as strong 
$\chi  \chi^+_1$  
coannihilations. For a non negligible higgsino component, the 
 neutralino annihilates into gauge bosons or $t\bar{t}$
and relic density is also decreased by $\chi\chi^+_1$ and
$\chi\chi^0_2$ coannihilations. 

In Fig.\ref{fig:mchi-Oh2-dGS} (resp. \ref{fig:mchi-Oh2-bplus}), decreasing
$\delta_{GS}$ (resp. $b_+$) in the moduli (resp. dilaton) dominated 
scenario corresponds to going from a dominant wino (resp. bino) 
to a dominant bino (resp. higgsino) 
%\footnote{In the dilaton dominated case the higgsino
%fraction depends also on $m_{3/2}$ which explains the spitting of the cloud of
%figure \ref{fig:mchi-Oh2-bplus}b)}.
 component. 
The theoretically predicted relic density can be checked
against cosmological observations. In particular, recent data 
on the cosmological microwave background (CMB) from the WMAP
satellite ~\cite{WMAP} constrain the dark matter relic density to be (at 
the 2$\sigma$ level) in the range $\Omega_{\rm CDM} {\rm h}^2 = 
0.1126^{+0.0161}_{-0.0181}.$
We indicate on Figs \ref{fig:mchi-Oh2-dGS}b) and \ref{fig:mchi-Oh2-bplus}b) 
the WMAP favoured range, keeping in mind that the
relic density calculation is very sensitive to SUSY mass and coupling 
uncertainties. In the following, we plot models in a generous 
range of relic density $0.03<\Omega h^2<0.3$, indicating also the WMAP constraint.

\subsection{Dark matter profiles}

 A crucial ingredient for the calculation of annihilation
fluxes is the density profile of dark matter, which is usually
parameterized as
\begin{equation}
  \rho(r)= \frac{\rho_0}{(r/R)^{\gamma}
  [1+(r/R)^{\alpha}]^{(\beta-\gamma)/\alpha}} \;\;.
\label{profile} 
\end{equation}
where $r$ is the galacto-centric coordinate, $R$ is a 
characteristic length and $\alpha, \beta$ and $\gamma$ are 
free parameters.
Unfortunately, large uncertainties are associated with such profiles,
especially in the innermost regions of galaxies, i.e. regions
where, in many cases,  most of the signal comes from. 

N-body simulations suggest the existence of 
'cuspy' profiles, following a power law $\rho(r)=r^{-\gamma}$ 
where $\gamma$ should be $\sim 1$ at small radii, although its exact 
value is under debate. Several groups tried to reproduce the initial 
results of Navarro, Frenk \& White \cite{Navarro:1996he},
who found $\gamma=1$, but reached different conclusions.
In Tab.~\ref{tab} we give the values of the parameters 
$(\alpha, \beta, \gamma)$ for some of the most widely used 
profile models, namely the Kravtsov et al. (Kra, \cite{Kravtsov:1997dp}
), 
Navarro, Frenk and White (NFW,  \cite{Navarro:1996he}), Moore et al. 
(Moore, \cite{Moore:1999gc}) and modified isothermal (Iso, e.g. 
\cite{Bergstrom:1997fj}) profiles. 
The most recent N-body simulations (Hayashi et al 2003) suggest 
that profiles do not approach power laws with a well defined 
index at very small radii. Profiles continue to become shallower,
i.e. the (negative) logarithmic slope becomes higher, when moving 
towards the centre. 

Furthermore, the presence of a $3.6\times 10^6$ 
solar masses black hole lying at the Galactic Center (see e.g.Ref.~\cite{schodel}) 
could possibly modify the profile of dark matter, that would accrete on it
producing a so-called 'spike'~\cite{Gondolo:1999ef}, leading to
an enhancement of the annihilation flux by several orders of 
magnitude. The prospects of indirect detection of dark matter 
in presence of such a spike have been discussed in Ref.~\cite{Bertone:2002je} 
(see ibid. for a discussion of the dynamical effects, recently 
proposed in literature, that could potentially destroy the spike). 
The observational situation is even more unclear.
Density profiles are usually reconstructed from the observation
of rotation curves of galaxies, in particular of low surface
brightness galaxies (LSB), that are thought to be dark matter dominated.
de Blok et al.~\cite{deblok} used this method to claim the inconsistency
of the observed 'flat' profiles, with the cuspy profiles 
predicted by n-body simulations. Other groups \cite{Swaters:2002rx,vandenBosch:1999ka} 
claimed instead that cuspy profiles are  
compatible with observations. 
Hayashi et al \cite{Hayashi:2003sj} compared the observational data with 
their numerical simulations (not with fitting formulae of 
their simulations) and found no significant discrepancy
in most cases. They attributed the remaining discrepancies
to the difference between circular velocities and gas rotation
speed in realistic triaxial halos. For more information on 
DM profiles see Ref.~\cite{Bertone:2004pz}. We assume in the following 
a NFW profile, and an observation angle of $10^{-3}$ sd,
the results for other profiles can be easily obtained 
using the numbers in Tab.1.

\begin{center}
\begin{table}
\centering
\begin{tabular}{|c|ccccc|}
\hline 
&$\alpha$&$\beta$&$\gamma$&R (kpc)&$\bar{J}\left( 10^{-3}\right)$ \\
\hline 
Kra& 2.0& 3.0&0.4 & 10.0 &$ 2.166 \times 10^1$ \\
NFW& 1.0& 3.0& 1.0& 20& $1.352 \times 10^3$\\
Moore& 1.5& 3.0& 1.5& 28.0 &$ 1.544 \times 10^5$ \\
Iso& 2.0& 2.0& 0& 3.5&$2.868 \times 10^1$\\ 
\hline 
\end{tabular}
\caption{Parameters of some widely used density profiles  models
and corresponding value of $\bar{J}(10^{-3})$.}
\label{tab}
\end{table}
\end{center}

\subsection{Flux of secondary particles}

Indirect detection of dark matter is based on the observation of 
secondary particles originating from dark matter annihilations in 
a cosmic storage area like the galactic halo. The study presented here is 
conceptually similar to the one 
in Ref.~\cite{Bertone:2002ms}, devoted to indirect detection
of Kaluza-Klein dark matter. 

The observed flux of secondary particle of species $i$, from the 
annihilation of dark matter particles of mass $M$ and annihilation
cross section $\sigma v$, from a direction $\psi$ and at energy $E$,
can be expressed as (e.g. Ref.~\cite{Bertone:2002ms, Jungman})
\begin{equation}
\Phi_i(\psi,E)=\sigma v \frac{dN_i}{dE} \frac{1}{4 \pi M^2}
\int_{\mbox{line of sight}}d\,s
\rho^2\left(r(s,\psi)\right)\label{flux}
\end{equation}
where $dN_i/dE$ is the spectrum of secondary particles per annihilation and
$r^2=s^2+R_0^2-2lR_0 \cos \psi$,
with $R_0\sim 8.5 \mathrm{kpc}$ is the solar distance to the galactic center.

It is customary (see \cite{Bergstrom:1997fj}), in order to separate 
the factors depending on astrophysics from those depending only 
on particle physics, to introduce the quantity $J(\psi)$
\begin{equation}
J\left(\psi\right) = \frac{1} {8.5\, \rm{kpc}} 
\left(\frac{1}{0.3\, \mbox{\small{GeV/cm}}^3}\right)^2
\int_{\mbox{\small{line of sight}}}d\,s\rho^2\left(r(s,\psi)\right)\,.
\label{gei}
\end{equation}
We then define $\bar{J}(\Delta\Omega)$ as  the average of $J(\psi)$ over
a spherical region of solid angle $\Delta\Omega$, centered on $\psi=0$.
The values of $\bar{J}(\Delta\Omega=10^{-3})$ are shown in the
last column of  Tab.~\ref{tab} for the corresponding density profiles.
In what follows, we assume a NFW profile : results corresponding to the other profiles of Table 1) can be deduced by a rescaling of $\bar{J}$

WIMP annihilation in the Galactic Halo produce a flux of gamma--rays, with 
either a continuum (coming from $\pi^0$ decay after the shower from hadronization
of the neutralino annihilation products (quarks or gauge/higgs bosons)) or a monochromatic distribution (produced from one--loop 
annihilation into $\gamma \gamma$ or $Z \gamma$). Monochromatic
processes are loop suppressed, and contribute little to the $\gamma$ spectrum.
But, as the neutralino can be considered at rest, the monochromatic ray is at an energy equal to the neutralino mass.
This could be one of the most promising signals for the discovery of supersymmetry in dark matter searches.

At the same time, annihilating neutralinos can also generate positron fluxes.
Cosmic--ray electrons and positrons interact with the interstellar medium
 through synchrotron radiation and inverse Compton scattering. The observed
 $e^{\pm}$ flux is dominated by primary electrons from acceleration sites,
however, 10 percent of the total flux is made of secondary electrons and
 positrons produced (in equal number) by the interaction of the primaries
 with the interstellar medium.

 Data from $HEAT$ (High Energy Antimatter Telescope) suggest
 the existence of a ``bump'' in the positron flux around 10 GeV
 \cite{Barwick:1997ig}. Halos of neutralinos could be a new source of positron
 explaining this excess, as there is no standard mechanism that would produce
 a signal at such a high energy. However, it is difficult to reproduce the
normalization of the observed positron fraction even assuming 
that fluxes are ``boosted'', e.g. because of the presence of dense
DM substructures in the solar neighborhood (see the discussion and references
in Ref.~\cite{Bertone:2004pz}).
Moreover, these secondary $e^{\pm}$ could propagate in the galactic magnetic
 field, generating synchrotron radiation (see below). 
  
\subsection{Experiments}

Up to now, detectors have explored energy ranges above $\sim 300$ GeV
(ground--based telescopes like Whipple or Cangaroo) or below $\sim 20$ GeV 
(EGRET). But the region which might turn to be the most promising
one for neutralino physics is the intermediate region
($30 ~\mathrm{GeV}<~E_{\gamma}~<300 ~\mathrm{GeV}$) that will be explored
by the next generation of detectors. In this paper we focus on
the complementarity of two type of telescope : a ground--based Atmospheric
\v{C}erenkov Telescope ({\it  High Energy Stereoscopic System}, HESS) and
a satellite experiment ({\it Gamma--Ray Large Area Space Telescope}, GLAST).

 Ground--based and
satellite experiments are complementary sources of data for supersymmetric
dark matter searches. While satellite experiments allow for a lower
energy threshold (less than 1 GeV for GLAST), a better energy resolution and 
a longer exposure time, the small effective area of the order of the square meter
limits  the sensitivity to high energy photons. On the contrary, the
large effective areas ($\sim 0.1$ ${\rm km^2}$ for HESS) of ground-based
Cerenkov telescopes permit the measurement
of very high energy fluxes with a higher energy threshold 
($E_{\gamma}> 60$ GeV for HESS)

\section{Analysis}

Gamma ray fluxes strongly depend on the phenomenology of the neutralino.
On the other hand, the nature of the neutralino is determined by the 
fundamental parameters of the models at high scale. We first analyze 
the dependence of neutralino annihilation on the main parameters of moduli
 dominated scenario ($m_{3/2}$, $\delta_{GS}$, $\tan \beta$) or dilaton
 dominated scenario ($m_{3/2}$, $b_+$) and then
the resulting gamma flux dependence on these parameters. We then compare
the predictions of our models with the sensitivities of present and future gamma
ray experiments. All calculations are achieved assuming an intermediate NFW 
profile: the results for other distributions can be easily deduced from a
rescaling by using the $\bar{J}$ values in the table \ref{tab}.

\subsection{Neutralino annihilation}

\begin{figure}[t]
\begin{center}
 \begin{tabular}{ccc}
\includegraphics[width=0.26\textwidth]{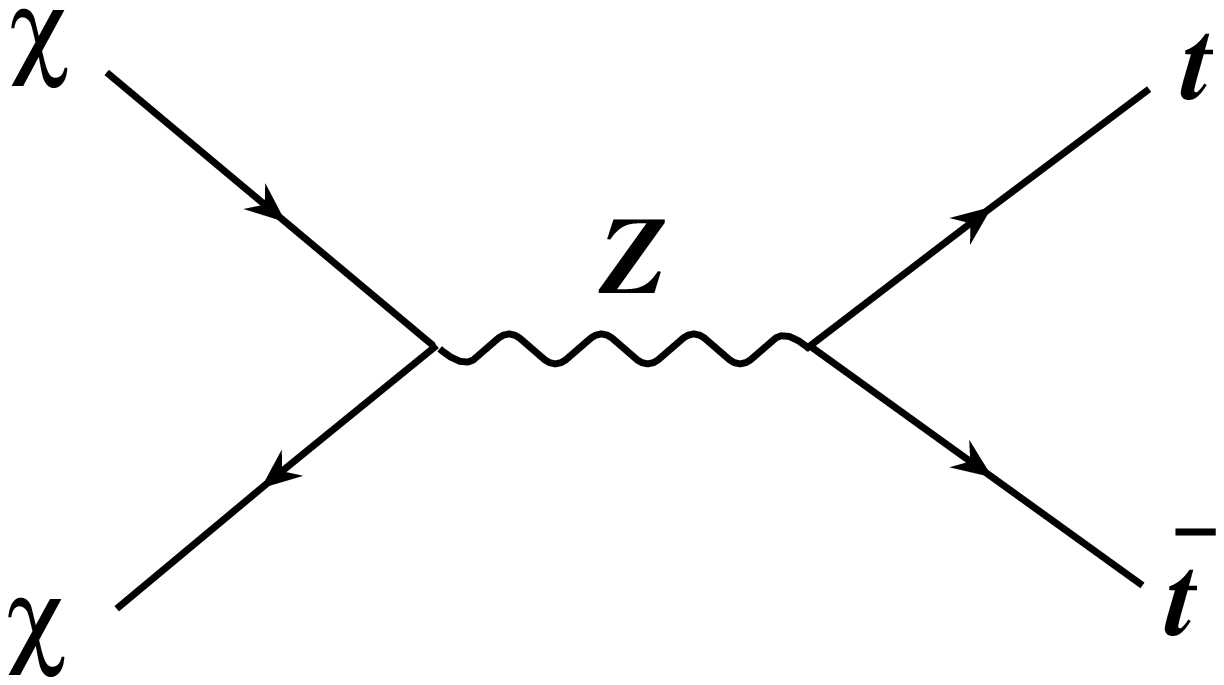}&
\includegraphics[width=0.26\textwidth]{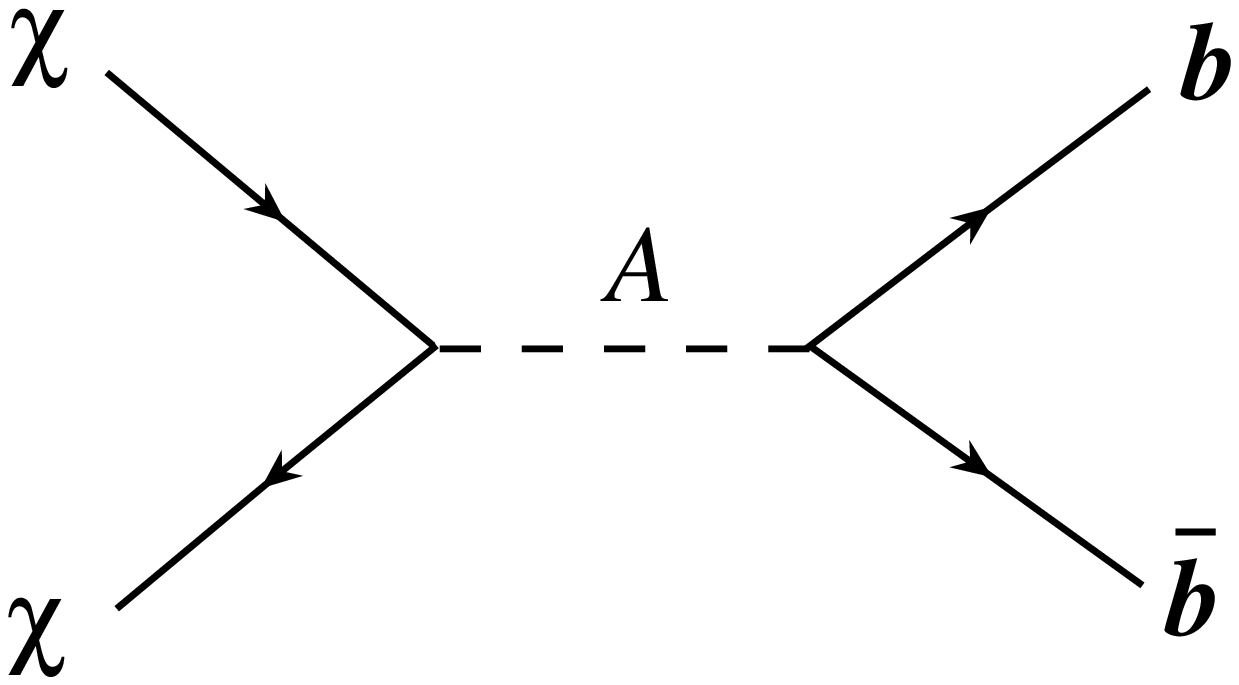}&
\includegraphics[width=0.17\textwidth]{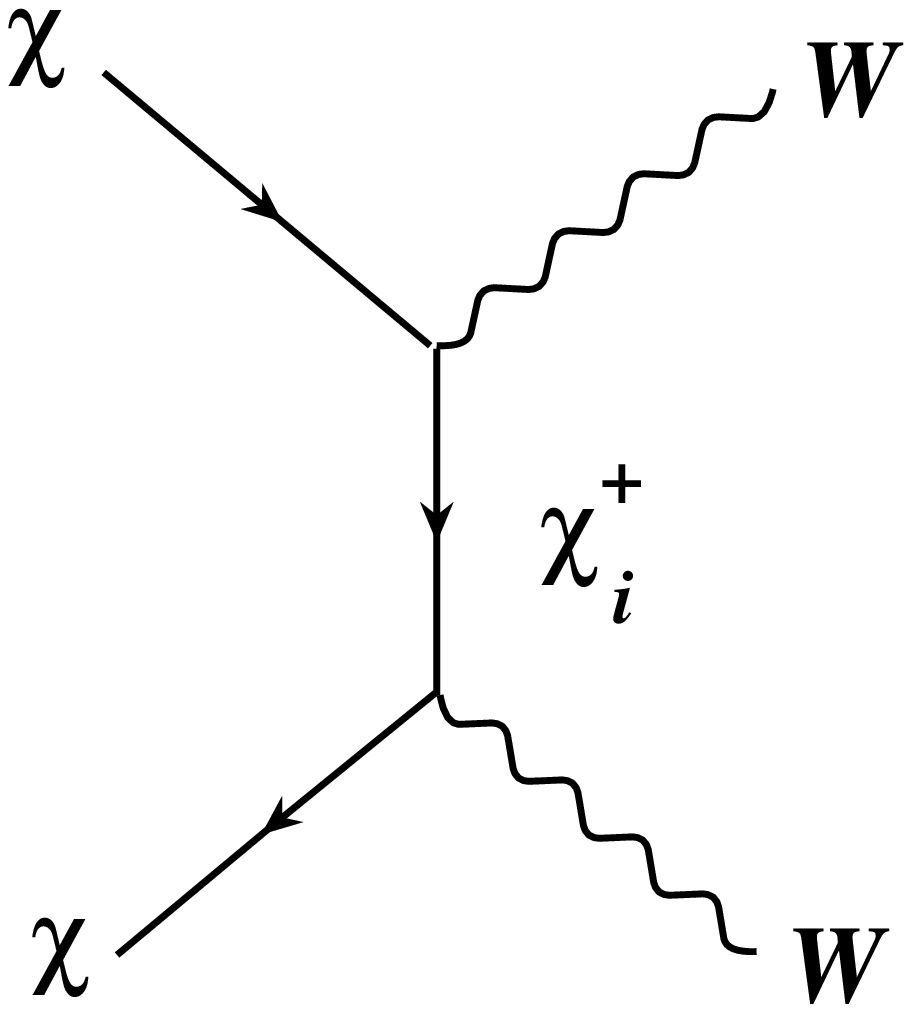}\\
 a) & b) & c)
\end{tabular}
 \caption{{\footnotesize {\bf The dominant Feynman graph}
 contributing for the process $\chi ~ \chi \rightarrow \gamma ~ \gamma$.}}  
\label{fig:feynman}
   \end{center}
\end{figure}

The main neutralino annihilation processes are $\chi \chi
\stackrel{Z}{\rightarrow} t\bar{t}$ ($\sigma_{t\overline{t}} 
\propto [z^2_{\chi3(4)}]^2$), $\chi
\chi \stackrel{\chi^+}{\rightarrow}W^+W^-$  ($\sigma_{WW} \propto
[z_{\chi3(4)}V_{12}]^2$ and/or $ [z_{\chi2}V_{11}]^2$ where $V_{11(2)}$ is
 the wino
(higgsino) component of the exchanged chargino) and  $\chi \chi
 \stackrel{A}{\rightarrow}b \bar{b}$ 
($\sigma_{b \overline{b}} \propto [z_{\chi1(2)}z_{\chi3(4)}]^2$). 
 Thus, both neutralino wino and
higgsino components strongly enhance annihilation. The processes are shown in Figs. \ref{fig:feynman}. The opening of the
different channels obviously depends on kinematics ($m_{\chi}$ versus
$m_b,m_W, m_t$). For a dominant bino LSP ($M_1 << M_2, ~ \mu$), the s--channel 
$A$ exchange (Fig. \ref{fig:feynman}d)  is the only one present. This
process is dominating for low $m_A$ especially
when $m_{\chi}$ approaches $m_A/2$ (A--pole) and is thus favored for high
values of $\tan{\beta}$ thanks to the $b \overline{b}$ coupling
to the pseudoscalar. When kinematically allowed, annihilation into
$t\overline{t}$  final state by $Z$ exchange can be also open despite the 
coupling suppression. For a wino--like neutralino ($M_2 << M_1, ~ \mu$), 
the
annihilation into $WW$ final state through $\chi_1^+$ exchange
(Fig. \ref{fig:feynman}c) is dominant thanks to the 
$\chi \chi^+_1 W$ coupling $and$
its propagator factor ($m_{\chi}~\sim M_2 ~\sim m_{\chi^+}$). For an
higgsino--like neutralino ($\mu << M_1, ~ M_2$), when $m_{\chi}>m_t$, the $Z$
exchange ($t\bar{t}$ final state) is enhanced through the $\chi \chi Z$
coupling.  But for even lower values of
$\mu$ ({\it i.e} neutralino mass), $t \overline{t}$ final state is
kinematically closed, and the neutralinos self annihilate into $W^+W^-$ or $b
\overline{b}$ (when $m_{\chi}<m_W$) final states.

The differential energy spectrum of the photon depends a lot on the primary
product of the neutralino annihilation. The gamma--ray spectrum was simulated
in \cite{Ullio3} and fitted well with the function 
$dN_{\gamma}/dx=ae^{-bx}/x^{1.5}$, where $x=E_{\gamma}/m_{\chi}$ and 
$(a,b)=(0.73,7.76)$ for $WW$ and $ZZ$, $(1.0,10.7)$ for $b\overline{b}$,
$(1.1,15.1)$ for $t\overline{t}$, and $(0.95,6.5)$ for $u\overline{u}$.
As we will see later on, final states with gauge boson will produce the harder spectrum.

%======================================================

%==============================================================================
\begin{figure}[t]
\begin{center}
 \includegraphics[width=0.8\textwidth]{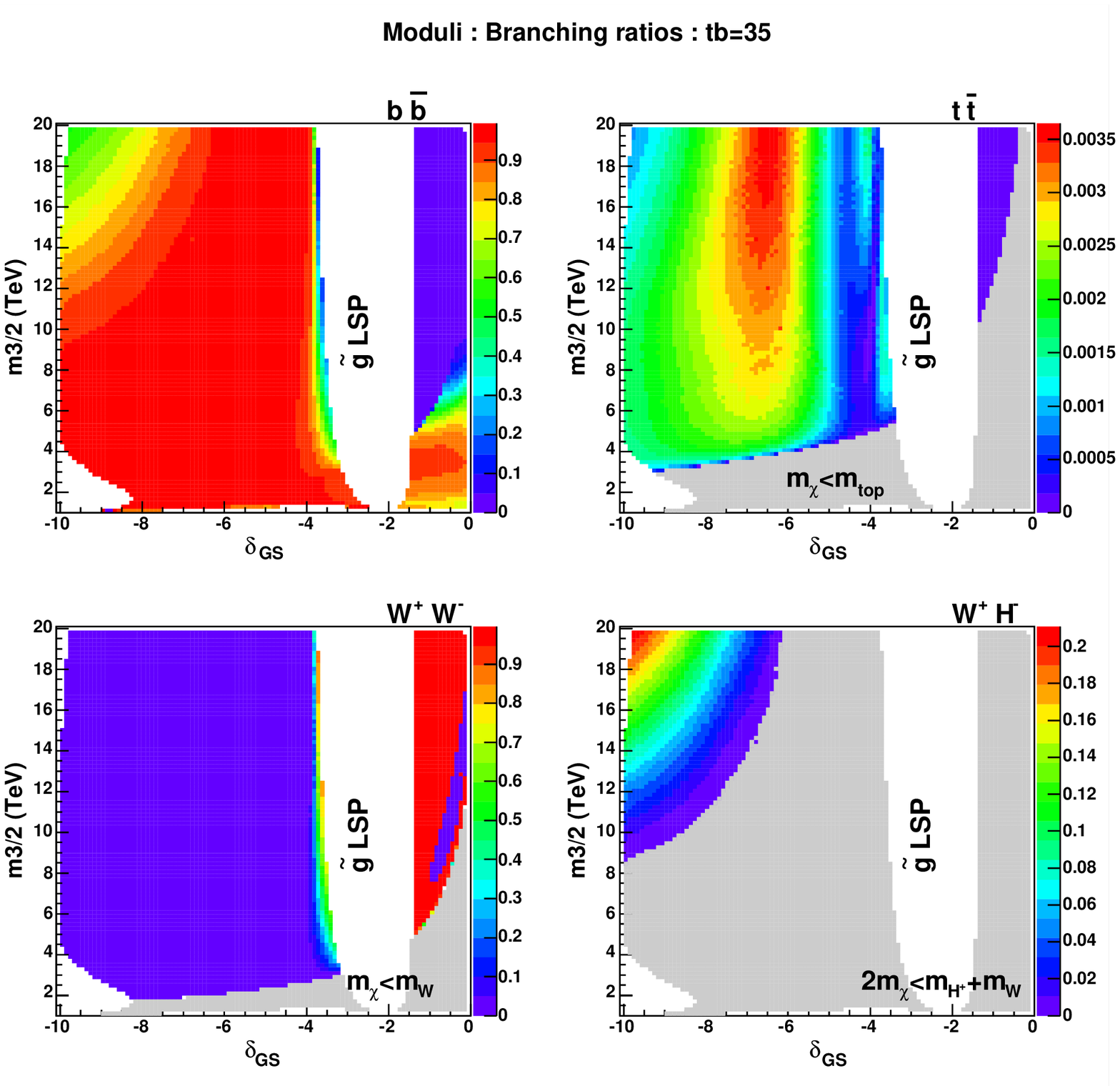}
\caption{\small Moduli domination regime: dominant annihilation branching ratios in the
  ($\delta_{GS},m_{3/2}$) plane for $\tan{\beta}=35,~t=0.25,~p=0$. Regions with gluino LSP are indicated. We
  also show in grey the kinematically forbidden region for
  each channel.}
\label{fig:BR-modtb35}
\end{center}
\end{figure}

%==============================================================================

%============ FIGURE 3ab : Gamma integrated Moduli dGS tbeta=5 and 35 ==========
\begin{figure}[h!]
\begin{center}
\begin{tabular}{cc}
 \includegraphics[width=0.45\textwidth]{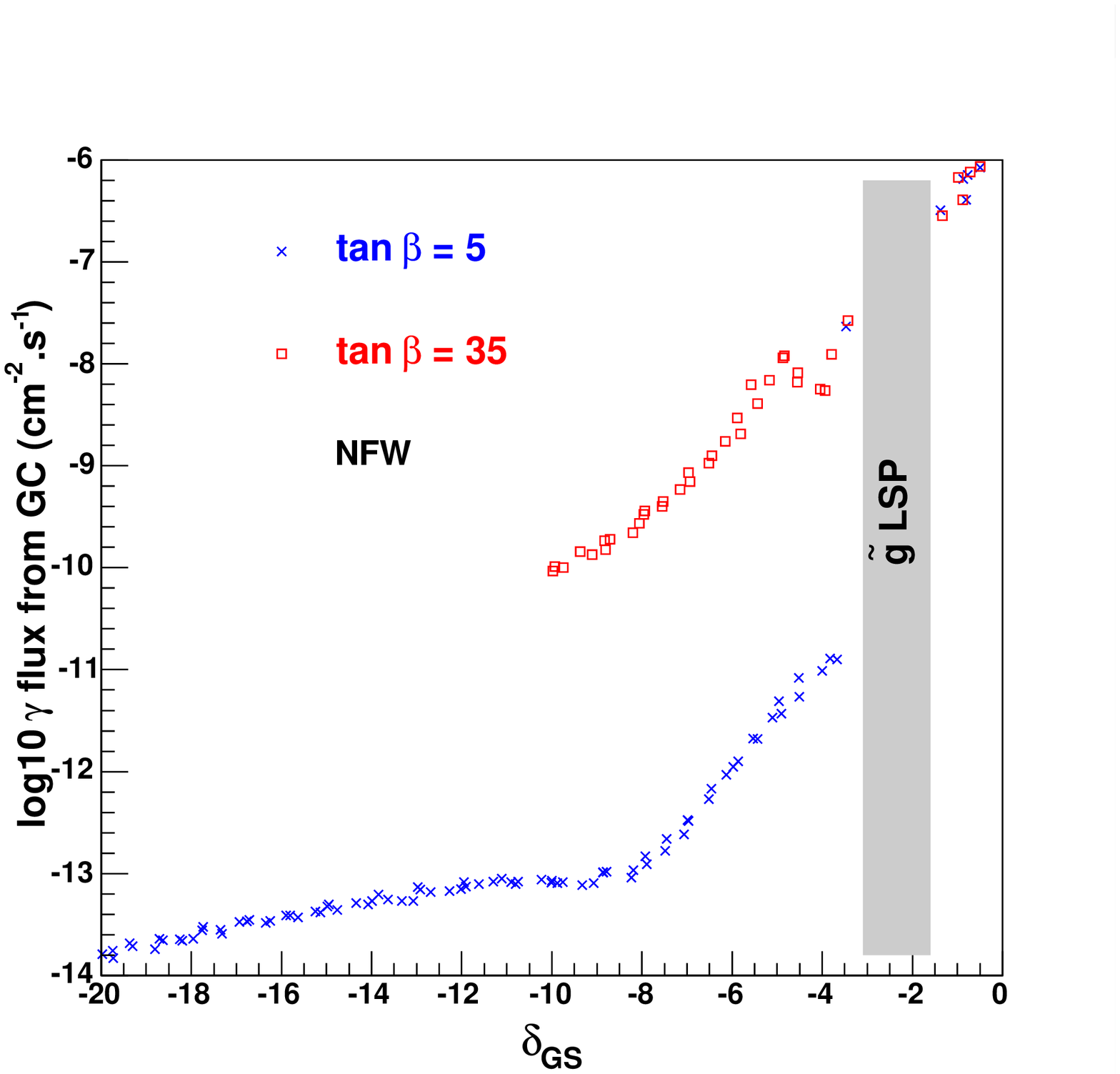}&
\includegraphics[width=0.45\textwidth]{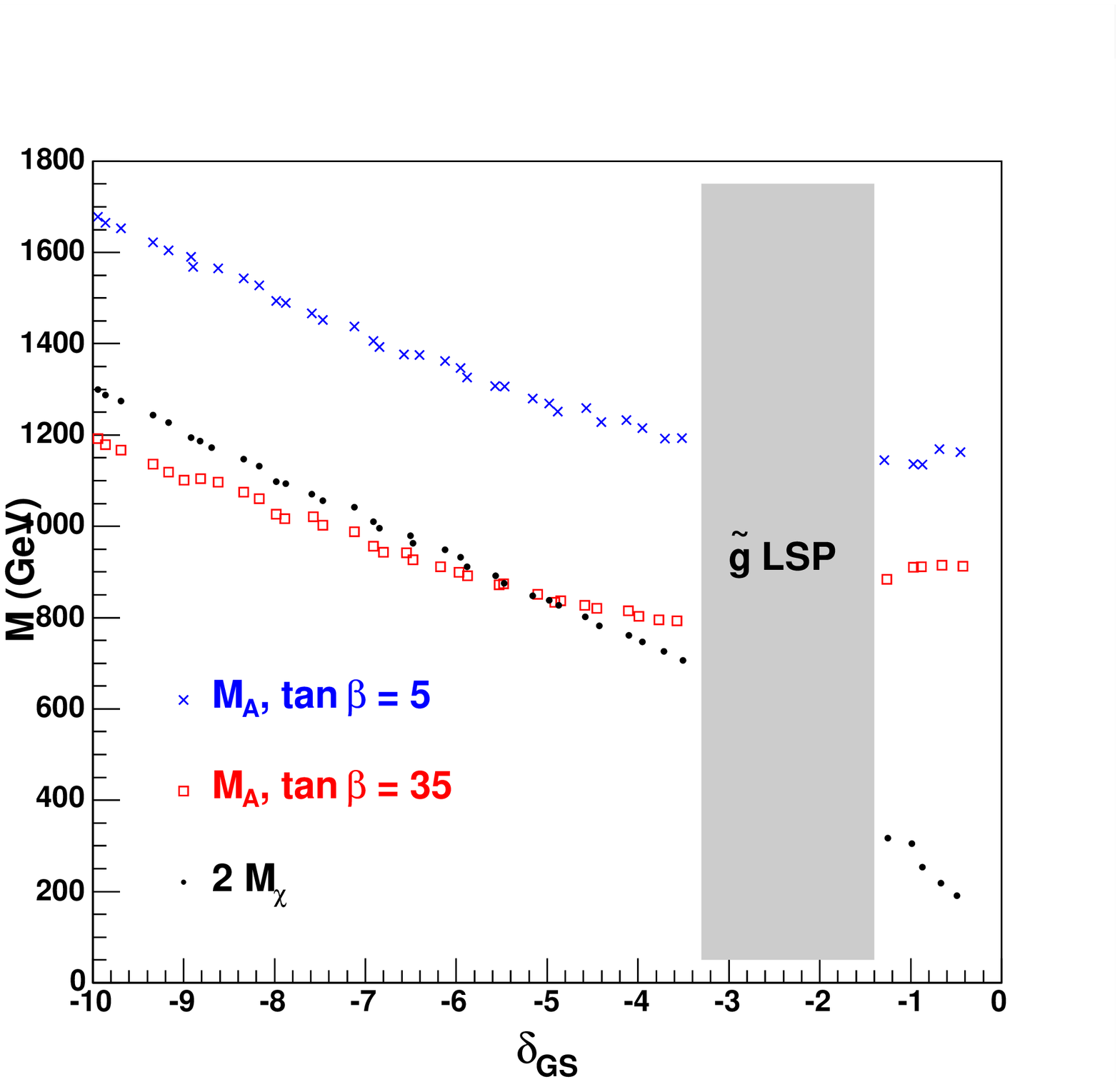}\\
a) & b)
\end{tabular}
          \caption{{\footnotesize Moduli-dominated scenario. a) The integrated gamma flux ($E_{\gamma}>1$
              GeV) for a NFW profile in the moduli parameter space as a function of the Green Schwarz counterterm $\delta_{GS}$, 
          for $m_{3/2} = 10\  {\rm TeV}, t=0.25,~p=0$, tan $\beta$ $=5$ (blue crosses) 
          and $35$ (red boxes) and b) the relevant mass spectrum of 
 as a function of $\delta_{GS}$ for the same values of tan $\beta$. See the text for the comments. }}
        \label{fig:gamintmoduli_dGS}
\end{center}
\end{figure}

%==============================================================================

\subsection{Gamma-ray flux in the moduli dominated scenario}

The integrated gamma flux gives a good understanding of the
phenomenological aspects of the moduli sector of the theory.

As an illustration, dominant annihilation branching ratios are represented as a function of
$\delta_{GS}$ and $m_{3/2}$ in Figs. \ref{fig:BR-modtb35} for
$\tan{\beta}=35$ (and in appendix Figs. \ref{fig:BR-modtb5} for 
$\tan{\beta}=5$) without any experimental cut on the parameter space.
We also show in Fig.~\ref{fig:gamintmoduli_dGS}a the integrated 
(from 1GeV to $m_{\chi}$) gamma flux for a NFW halo profile as a function of 
the Green--Schwarz counterterm 
$\delta_{\mathrm{GS}}$ for two values of tan$\beta$ (5,35) and
$m_{3/2}=10$ TeV. 
To complete our illustration, 
we have plotted in Fig. \ref{fig:gamintmoduli_dGS}b the mass spectrum $2
m_{\chi}$  and $M_A$ for the same set of parameters. 
The mass of the neutralino (increasing with $|\delta_{GS}|$, 
Eq. \ref{modsoftgaugi}) 
and its nature (from wino to bino with increasing $|\delta_{GS}|$) determine 
in a large part the integrated flux. Indeed the LSP is mainly bino--like for 
$-20 < \delta_{GS} < -3$
and wino--like\footnote{We can compute the 
analytical expression
of $\delta_{GS}$ where this change of nature appears by solving 
in Eq. \ref{modsoftscal}
$M_1^{GUT} = 2 M_2^{GUT}$ ($M_1^{LOW}=M_2^{LOW}$) that 
gives $\delta_{GS}=-\frac{46}{15} \sim -3$.} 
for $-3 < \delta_{GS} < 0$.

Fig. \ref{fig:gamintmoduli_dGS}a illustrates first the general increase 
of the flux
as a function of $\delta_{GS}$, due to the decrease of the neutralino mass 
(Fig. \ref{fig:gamintmoduli_dGS}b).
For $\delta_{GS}$ running from -10 to -4,
we observe a higher flux for higher tan$\beta$ because
 the A--pole channel is kinematically open for
$\delta_{\mathrm GS} \sim -5$ at $\tan \beta = 35$
(value of $\delta_{GS}$ that gives $2m_{\chi}>M_A$). This process 
produces a large
amount of $b\overline{b}$ final states (Fig. \ref{fig:BR-modtb35}) thanks to 
the coupling $Ab\overline{b}$, 
proportional to tan$\beta$. The low tan$\beta$ (Fig. \ref{fig:BR-modtb5}) 
dominant channel is the
 $t\overline{t}$, proportional to 
$\frac{m_t m_{\chi}}{m_Z^2}$.
%Annihilation into $b\bar{b}$ and flux are higher for tan$\beta=35$ because of
%the $A$ exchange channel enhancement. 
For $-2 < \delta_{GS} < 0$, the neutralino is
entirely $wino$, leading to annihilation into $WW$ 
(Fig. \ref{fig:BR-modtb35}). Indeed, a crude approximation tell us
that the $WW$ cross section is proportional to $z_{\chi 2}^4 (\sim 1$ for
$\delta_{GS}=-1$) and the annihilation into $t \overline{t}$, 
proportional
to the square of the $\chi \chi Z$ coupling 
($\[z_{\chi 3}^2+z_{\chi 4}^2\]^2 \sim 10^{-5}$
for $\delta_{GS}= -4.$). This explains the 5 orders of magnitude observed 
on the Fig. \ref{fig:gamintmoduli_dGS} between the small and large values 
of $|\delta_{GS}|$ for $\tan{\beta}=35$.

For $\delta_{GS} \sim 0$ and small values of $m_{3/2}$, when the $WW$ channel
 is kinematically closed, dominant 
$\gamma$--ray contribution comes from the annihilation into $b\bar{b}$
through $Z$ and $A$ exchange (suppressed because proportional to $m_b$). For 
$\tan{\beta}=5$ and for
 higher $m_{3/2}$ values one can have $t\bar{t}$ or $gg$ annihilation channels
 which explains the change of slope of the flux with $\tan{\beta=5}$ on Fig. \ref{fig:gamintmoduli_dGS}.

\subsection {Gamma-ray flux in the dilaton dominated scenario}

In scenarios where the supersymmetry is spontaneously broken through
the $vev$  of the dilaton auxiliary field, the $\beta$ function $b_+$ of the
 first gaugino condensing group plays 
a role similar to the Green--Schwarz counterterm $|\delta_{GS}|$ in the
 moduli domination scenario. Since $b_+$  
does not depend on the gauge group indices of the Standard Model  
$SU(3) \times SU(2) \times U(1)$, it contributes universally to the gaugino mass 
breaking terms. 
Thus, the phenomenology of dilaton scenarios at large $b_+$ in the gaugino sector is similar
to the one of the moduli dominated models at large $|\delta_{GS}|$.
The scalar sector is responsible of most of the phenomenological differences.
Indeed, the scalar masses are non--universal and suppressed in 
moduli like scenario because of the loop anomaly factors suppression 
$\gamma_i \sim \frac{1}{16 \pi^2}$ (Eq. \ref{modsoftscal}), 
whereas they are universal and large in a dilaton dominated scenario 
($m_0 \sim m_{3/2}$ Eq. \ref{dilatsoftscal}). The main consequences are 
 heavier pseudoscalar\footnote{We are in the so called decoupling regime.} 
 $A$ and
$M_{H_u}^2$ breaking term at GUT scale, that drives the $\mu$ parameter to
 lower values, especially in high tan$\beta$ regimes, Eq.(\ref{mu}).

\begin{figure}[h!]
\begin{center}
\begin{tabular}{cc}
 \includegraphics[width=0.45\textwidth]{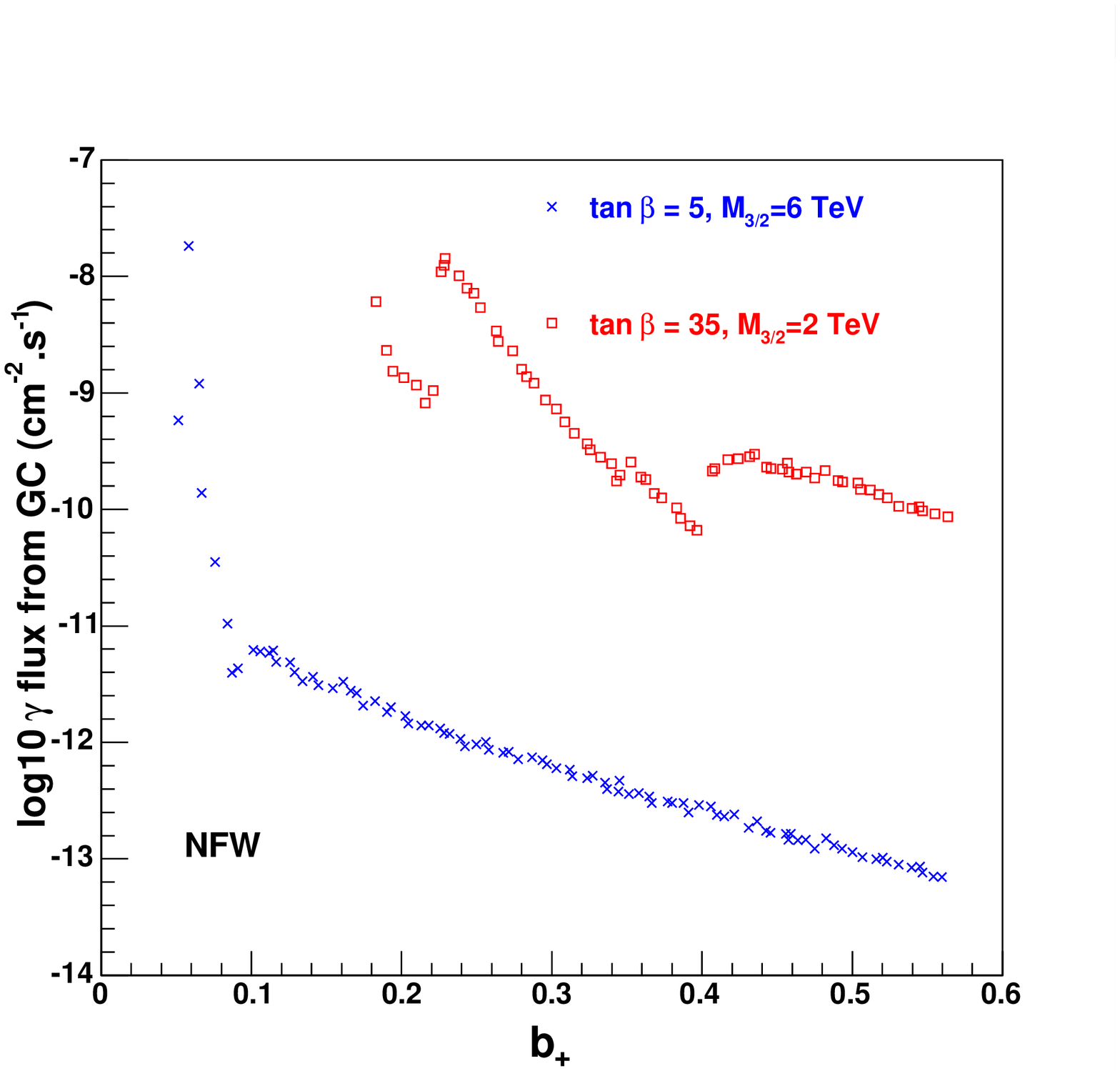}&
\includegraphics[width=0.45\textwidth]{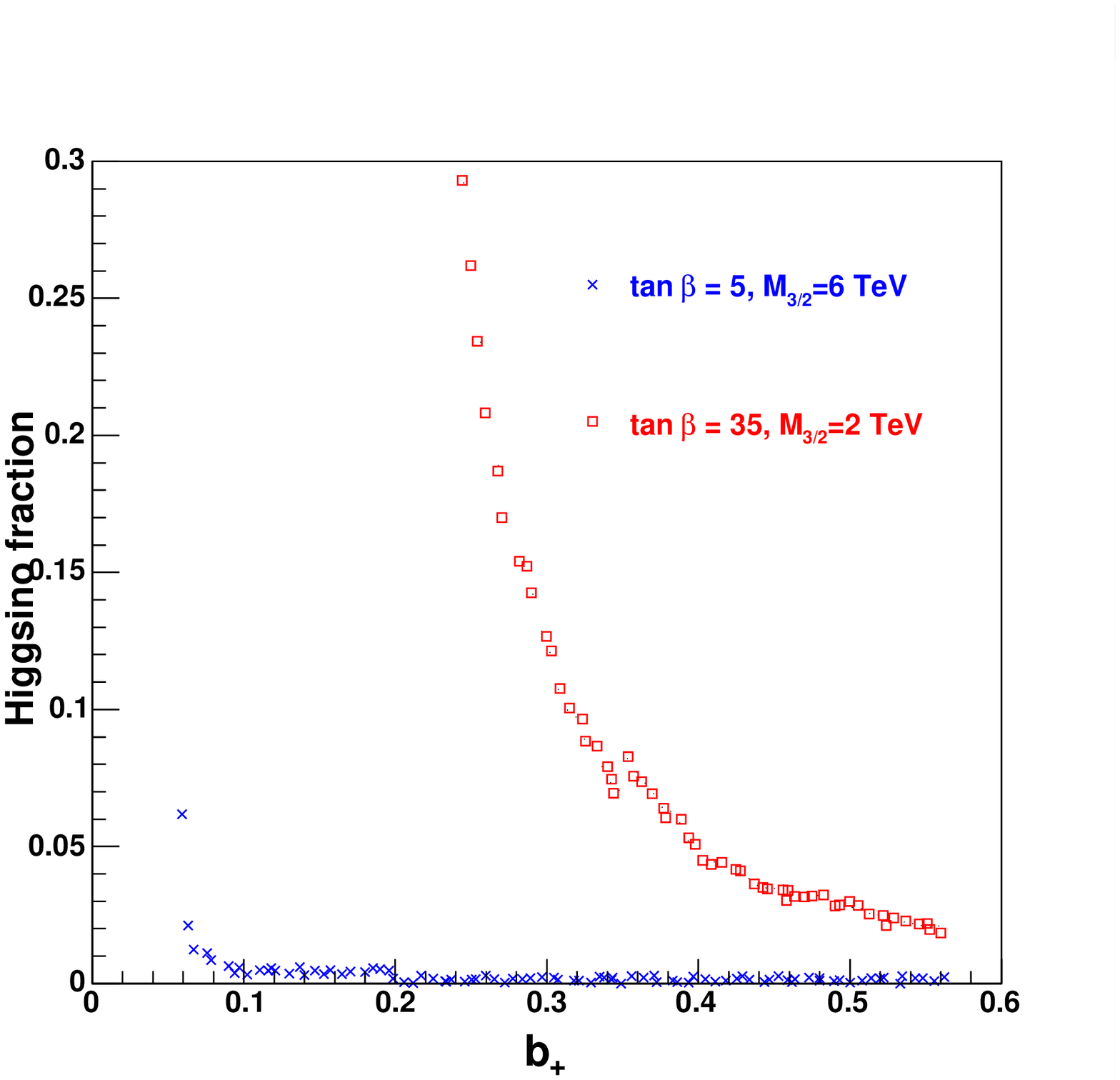}\\
a) & b)
\end{tabular}
\caption{{\footnotesize Dilaton-dominated scenario. a) The integrated gamma flux ($E_{\gamma}>1$
              GeV) for a NFW profile in the dilaton parameter space as a 
function of the beta  function $b_+$, for $t=0.25,~p=0$, tan $\beta$ $=5$ 
(blue crosses) and $35$ (red boxes) and b) the neutralino higgsino fraction as a
 function 
of $b_+$ for the same values of tan $\beta$. 
%Right : the mass spectrum for tan$\beta=35$
% as a function of $b_+$ for the same values of tan $\beta$. 
See the text for the comments. }}
        \label{fig:gamintdilaton_b+}
\end{center}
\end{figure}

As in the moduli dominated scenario, we show the dominant annihilation
 branching ratios, in the $(b_+, m_{3/2},)$ plane for 
$\tan{\beta}=35$ in Figs.\ref{fig:BR-diltb35} (and in appendix in 
Figs.\ref{fig:BR-diltb5} for $\tan{\beta}=5$) to highlight the different
 annihilation channel dependence on fundamental parameter.
We also show (in Fig. \ref{fig:gamintdilaton_b+}a) the integrated gamma flux
for a threshold $E_{\gamma}>1$ GeV, from the Galactic Center for a NFW profile, as a function 
of $b_+$, for two values of tan$\beta$ (5 and 35) and the higgsino fraction
of the lightest neutralino in Fig. \ref{fig:gamintdilaton_b+}b.

The effect of $b_+$ on the neutralino
nature is complex and indirect. 
It acts on $M_3$ through the $F^S$ contribution with a different sign
 contribution than the superconformal contribution proportional to
$b_3$ (Eq. (\ref{dilatsoftgaugi}).
Thus, lowering $b_+$ acts through the $M_3$ dependence of the RGE on 
$M_{H_u}^2$.
Lower $b_+$ means lower $M_3$, Eq.(\ref{dilatsoftgaugi}) leading to higher 
$M_{H_u}^2$ through the RGE implying a lower $\mu$ parameter, Eq.(\ref{mu}):
neutralino is mainly higgsino for $b_+ < 0.1(0.3)$ and tan$\beta$=5
(35) as we can see in Fig. \ref{fig:gamintdilaton_b+}b. Though quite low, 
its wino component is not completely negligible.

%==============================================================================

\begin{figure}[t]
\begin{center}
 \includegraphics[width=0.8\textwidth]{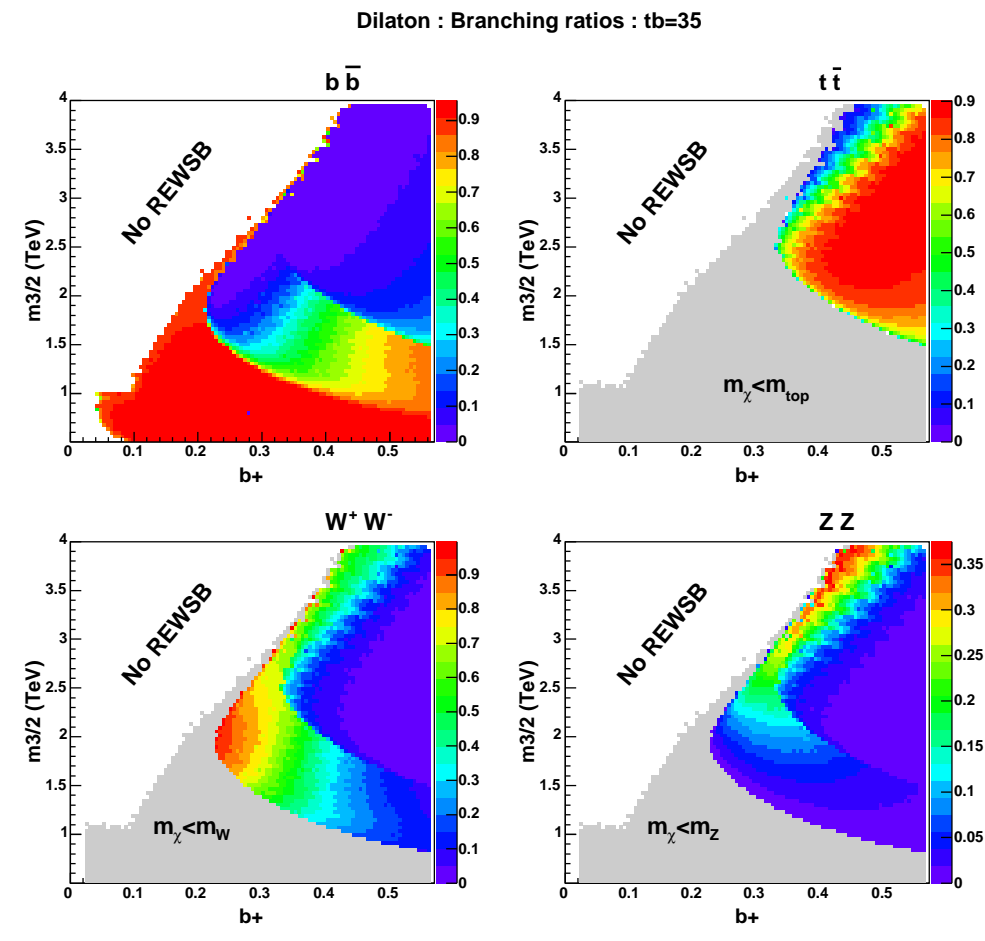}
\caption{\small Dilaton domination regime: dominant annihilation branching ratios in the
  ($b_{+},m_{3/2}$) plane for $\tan{\beta}=35$, $t=0.25,~p=0$. Regions where radiative
  electroweak symmetry breaking can not occur are indicated (No REWSB). We
  also show in grey the kinematically forbidden region for
  each channel.}
\label{fig:BR-diltb35}
\end{center}
\end{figure}

%==============================================================================

%============ FIGURE 4ab : Gamma integrated Dilaton b+ tbeta=5 and 35 ==========

%==============================================================================

For a fixed value of $m_{3/2}$, decreasing $b_+$ lead to a higher higgsino
fraction favoring $WW$, $ZZ$ channels and the $t\bar{t}$ one when
kinematically open ({\it i.e} for high $m_{3/2}$ values) compared to the
$b\bar{b}$ channel. For low $m_{3/2}$ values the wino component also favors
the $WW$ channel. When $m_{\chi}<m_W$, annihilation into $b\bar{b}$ is
completely dominant, the other processes being kinematically closed. 
This is illustrated on Figs. \ref{fig:BR-diltb35} for $\tan{\beta}=35$ and 
in appendix Fig. \ref{fig:BR-diltb5} for  $\tan{\beta}=5$.
For a given $m_{3/2}$, higher values of tan$\beta$ give better fluxes 
because it enhances the 
higgsino fraction of the neutralino through Eq.(\ref{mu}).
The neutralino mass increasing with $b_+$ (Fig.\ref{fig:mchi-Oh2-bplus}a)
explains the general evolution of the gamma flux 
($\propto \frac{1}{m_{\chi}^2} \sim \frac{1}{M_1^2}$) 
(Fig. \ref{fig:gamintdilaton_b+}a). 
The gamma flux follows also the higgsino fraction 
(Fig. \ref{fig:gamintdilaton_b+}b) which enhances the annihilation. The
different peaks correspond to changes in hierarchy between $M_1,\mu, M_W$
 in the neutralino mass matrix Eq. \ref{eq:matchi} exchanging the annihilation
 processes (Fig. \ref{fig:BR-diltb35}). The different flux slopes (Fig.\ref{fig:gamintdilaton_b+}) correspond
 to the different spectrum shape of each neutralino annihilation final
 states (Fig.\ref{fig:BR-diltb35}). 

\subsection{Comparison of models in view of the experimental sensitivities}

The phenomenology of the two SUSY breaking
 mediation sectors strongly depends on the fundamental parameter space 
of the models at GUT scale. 
It is then interesting to look back at the prospect of discovering
 gamma fluxes as a function of a physical parameter, the neutralino mass
 ($m_{\chi}$), and to compare
both SUSY breaking scenarios with mSUGRA. For that purpose,
we have computed the total gamma flux as a function of $m_{\chi}$ for three
 values of tan$\beta$, scanning on the two other 
main parameters of each model ($\[m_{3/2},\delta_{GS}\]$ in the moduli 
dominated case and 
$\[m_{3/2},b_+\]$ in the dilaton dominated one). We have applied the 
experimental cuts
 coming from LEPII constraints on SUSY spectrum, higher order processes and 
neutralino relic density favored regions (the details are developed
in appendix C).

\begin{figure}[h!]
\begin{center}
\begin{tabular}{cc}
 \includegraphics[width=0.45\textwidth]{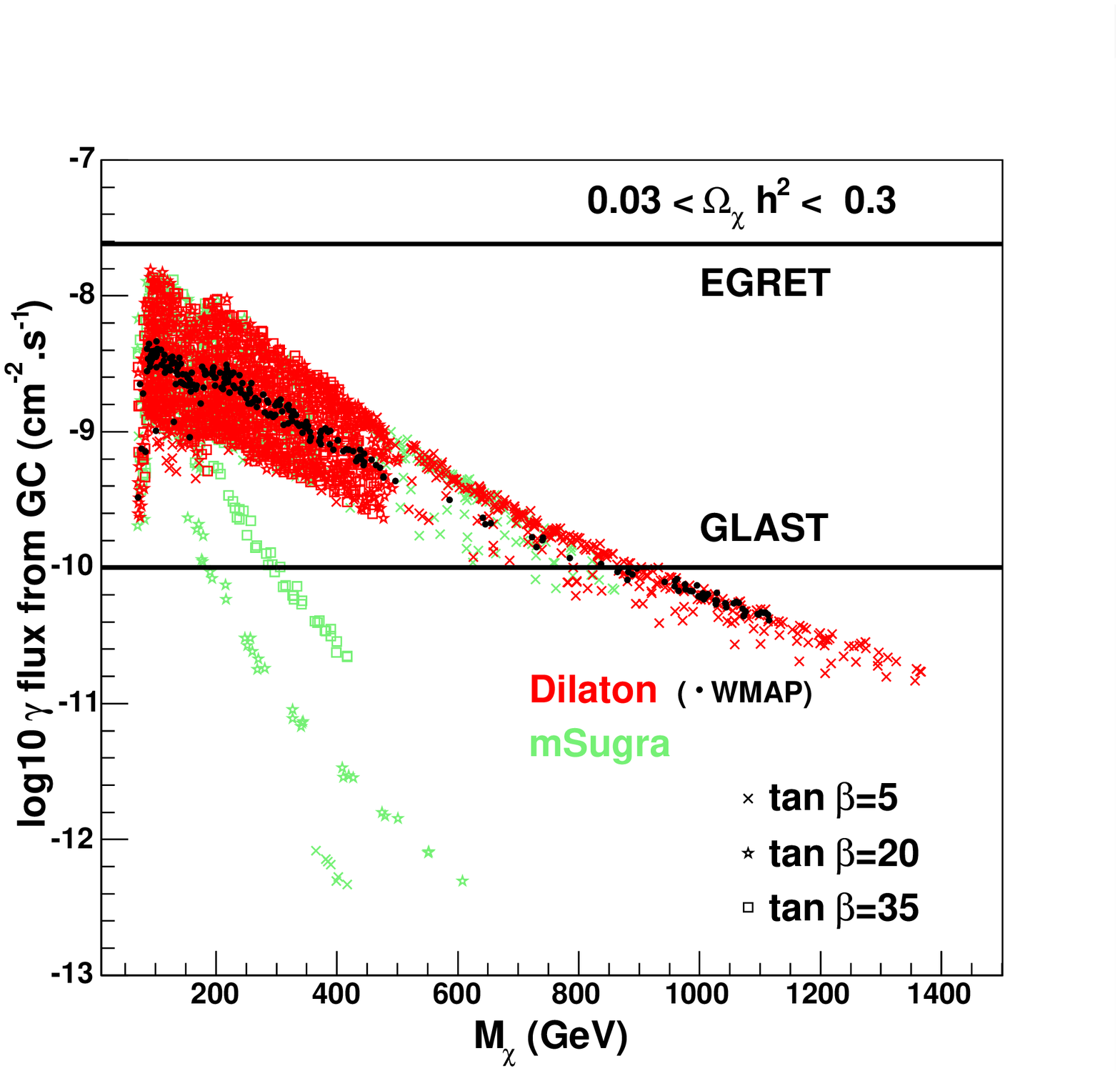}&
\includegraphics[width=0.45\textwidth]{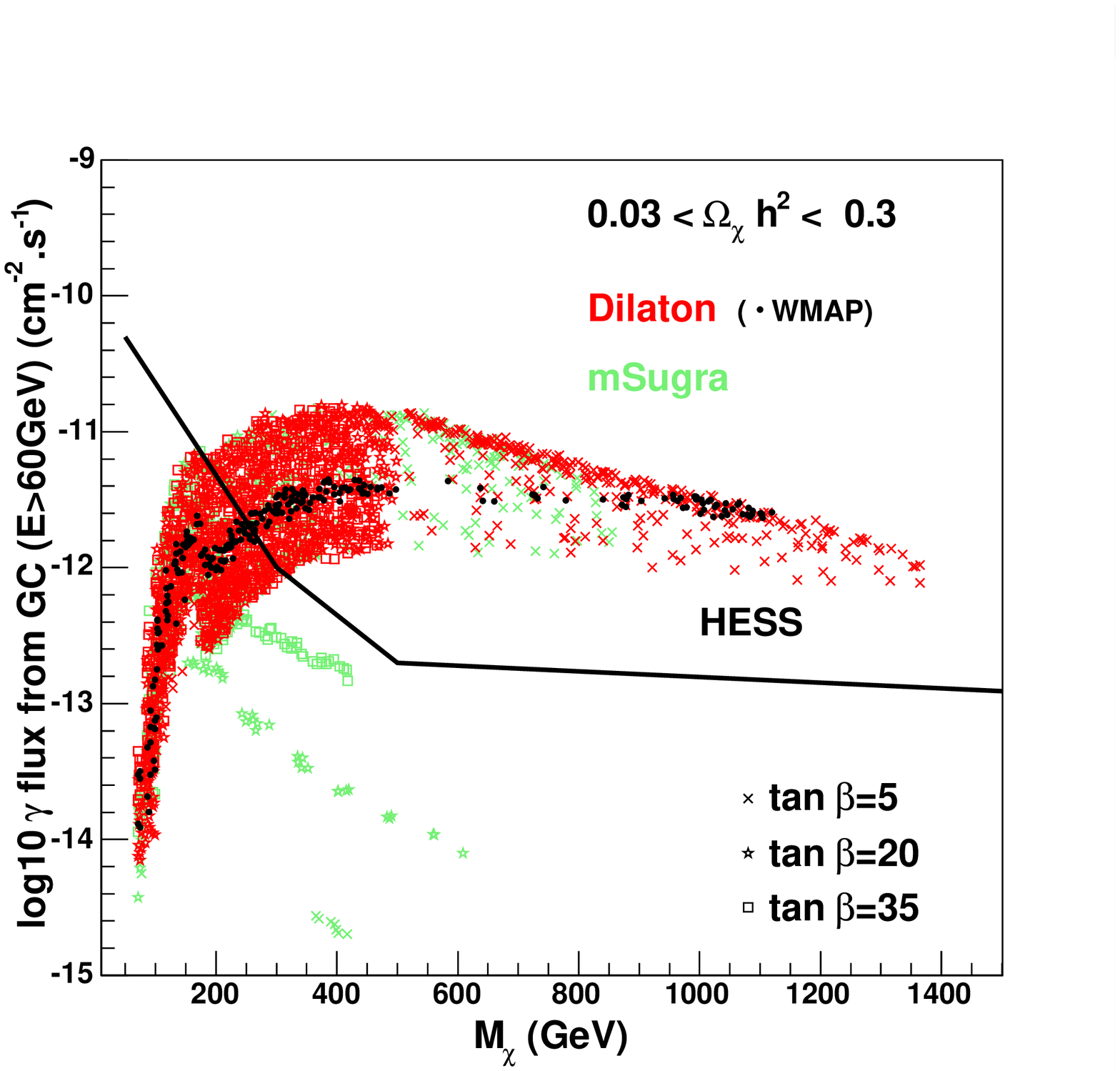}\\
a) & b)
\end{tabular}
\caption{{\footnotesize {\bf DILATON : The integrated flux of $\gamma$-ray} 
above 1 GeV in a) and  60 GeV in b) per centimeter square per second versus 
neutralino mass
for tan$\beta=5, 20$ and $35$ taking a NFW profile for the halo,
after a scan in the dilaton parameter space for $t=0.25,~p=0$. The points that 
violate the accelerator constraints
(see the appendix for more details) are not represented here. Sensitivities of 
present
 experiments and projects are represented in solid lines. We also show an
 ``equivalent'' mSUGRA cloud  (see discussion in appendix for the choice of
 parameter space).}}
\label{fig:fluxdil}
\end{center}
\end{figure}

 We have first plotted in Figs \ref{fig:fluxmod}a and \ref{fig:fluxdil}a 
the total gamma flux coming from the Galactic Center 
above 1 GeV as a function of the neutralino mass 
for a NFW profile and different values of $\tan \beta$ (5, 20, 35)
in the moduli and dilaton dominated scenarios.
We note that:
\begin{itemize}
\item For a $fixed$ value of tan $\beta$ and a given profile density, 
the regions covered by the two models do not overlap. In other words, if 
we obtain the value of $\tan \beta$ by another experiment, the
 measurement of the gamma-ray flux will allow to distinguish or even 
exclude SUSY breaking scenarios.
\item The flux is a decreasing function of the LSP mass, 
independently of the specific model adopted (see eq. \ref{flux}): 
a heavier neutralino has a lower number density in the halo,
and so a lower flux. Except some possible threshold effects, this is a 
general remark 
for any nature of indirect fluxes coming from neutralino annihilation at the 
Galactic Center. 
On the contrary, for lighter neutralino (dilaton--dominated scenario
 and mSugra) the threshold 
effect implies an increasing flux with $m_{\chi}$. This effect is evaded for 
$m_{\chi} \gtrsim m_{t}$, when the $t\bar{t}$ channel is open.
\item  Fluxes in the dilaton breaking scenario are higher than in the
 moduli scenario (except for high tan$\beta$ because of the $A$--pole
 contribution, see below). The highest flux in moduli dominated case is 
obtained  for low $|\delta_{\mathrm{GS}}|$
(Fig. \ref{fig:gamintmoduli_dGS}a),
where the neutralino is completely wino but does not have a sufficient relic
density to fulfill the dark matter content of the universe 
(see Fig.\ref{fig:mchi-Oh2-dGS}). Those points are excluded by our cut 
on $\Omega > 0.03$. The remaining ones
are situated in the near--zone "stau LSP" branch of Figs. 
\ref{fig:BR-modtb35} and \ref{fig:BR-modtb5}. 
\end{itemize}

\begin{figure}[h!]
\begin{center}
\begin{tabular}{cc}
 \includegraphics[width=0.45\textwidth]{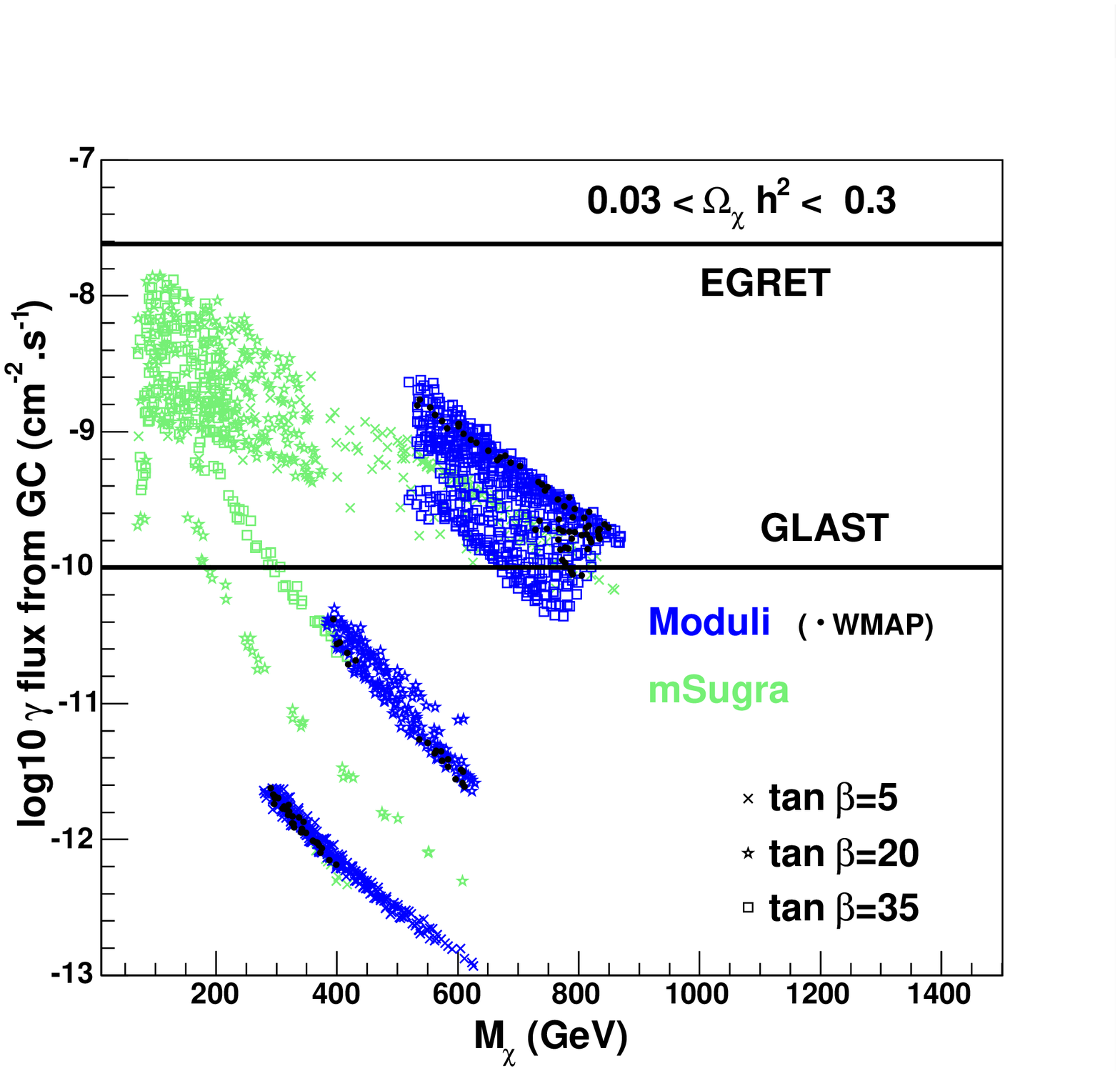}&
\includegraphics[width=0.45\textwidth]{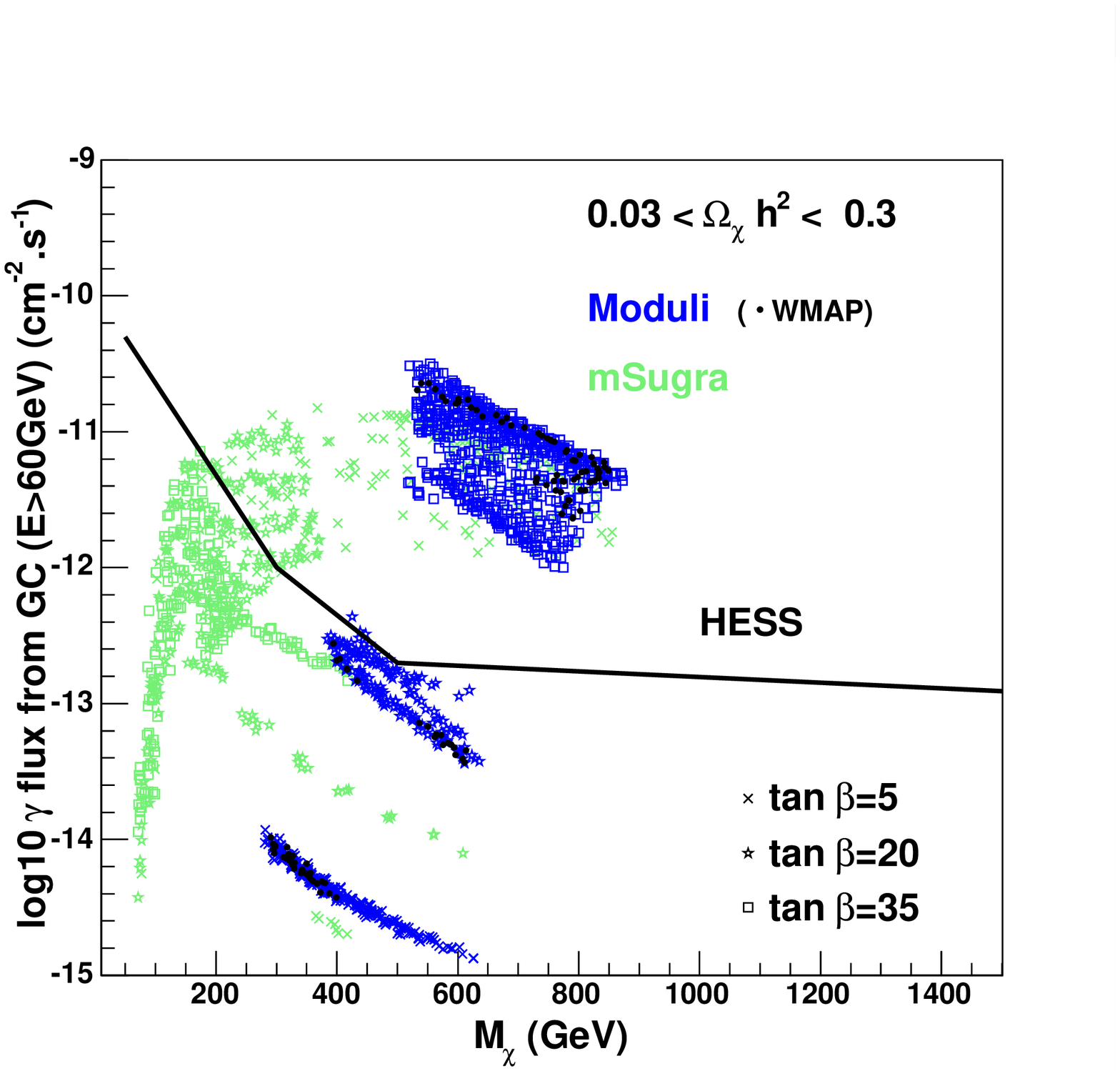}\\
a) & b)
\end{tabular}
\caption{{\footnotesize {\bf MODULI : The integrated flux of $\gamma$-ray} 
above 1 GeV in a) and  60 GeV in b) per centimeter square per second versus 
neutralino mass
for tan$\beta=5, 20$ and $35$ taking a NFW profile for the halo,
after a scan in the moduli parameter space for $t=0.25,~p=0$. The points that 
violate the accelerator constraints
(see the appendix C for more details) are not represented here. Sensitivities of 
present and future
 experiments are represented in solid lines. We also show an
 ``equivalent'' mSUGRA cloud  (see discussion in appendix B for the choice of
 parameter range).}}
\label{fig:fluxmod}
\end{center}
\end{figure}

\begin{figure}
\begin{center}
\begin{tabular}{cc}
 \includegraphics[width=0.45\textwidth]{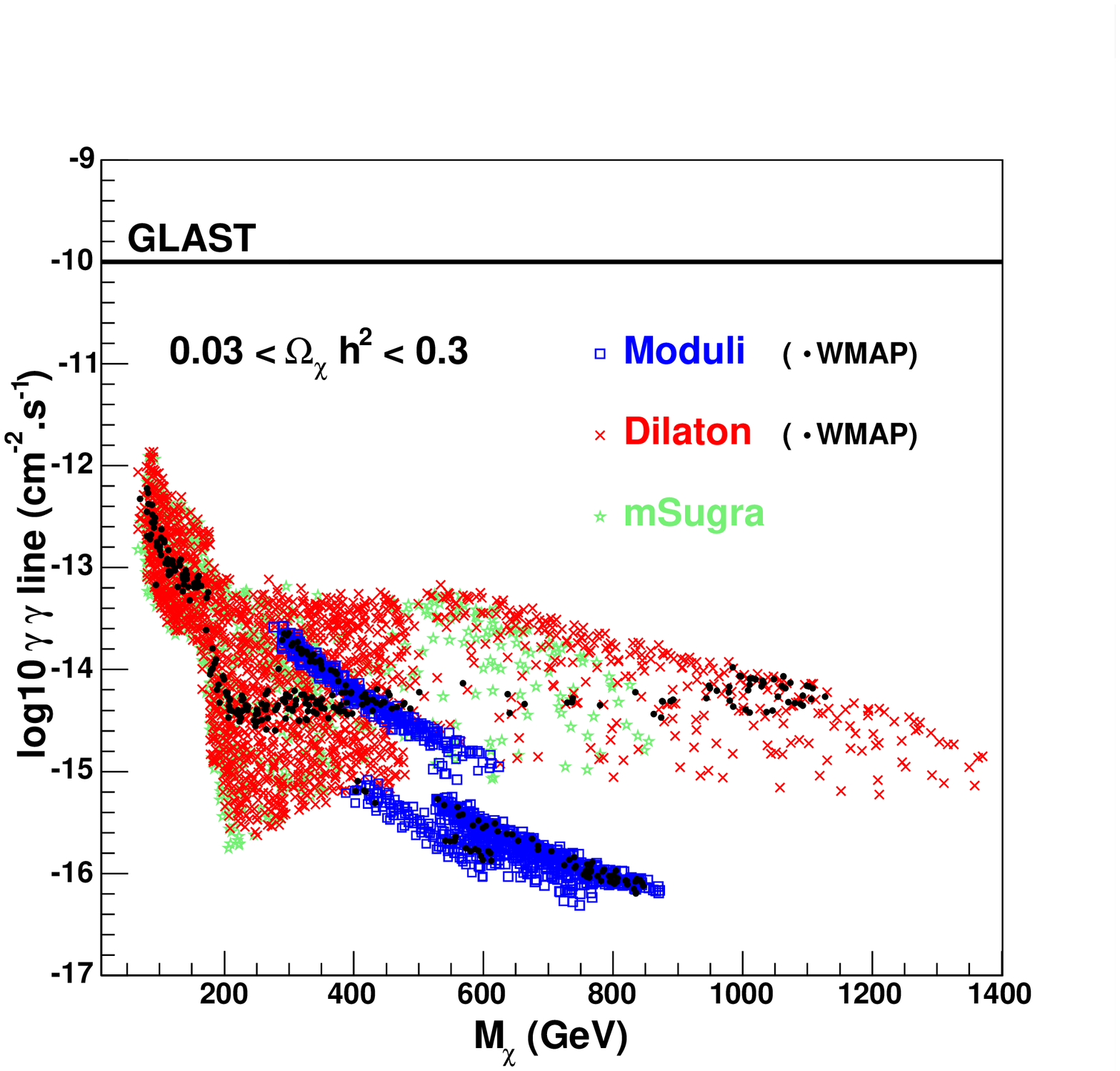}&
\includegraphics[width=0.45\textwidth]{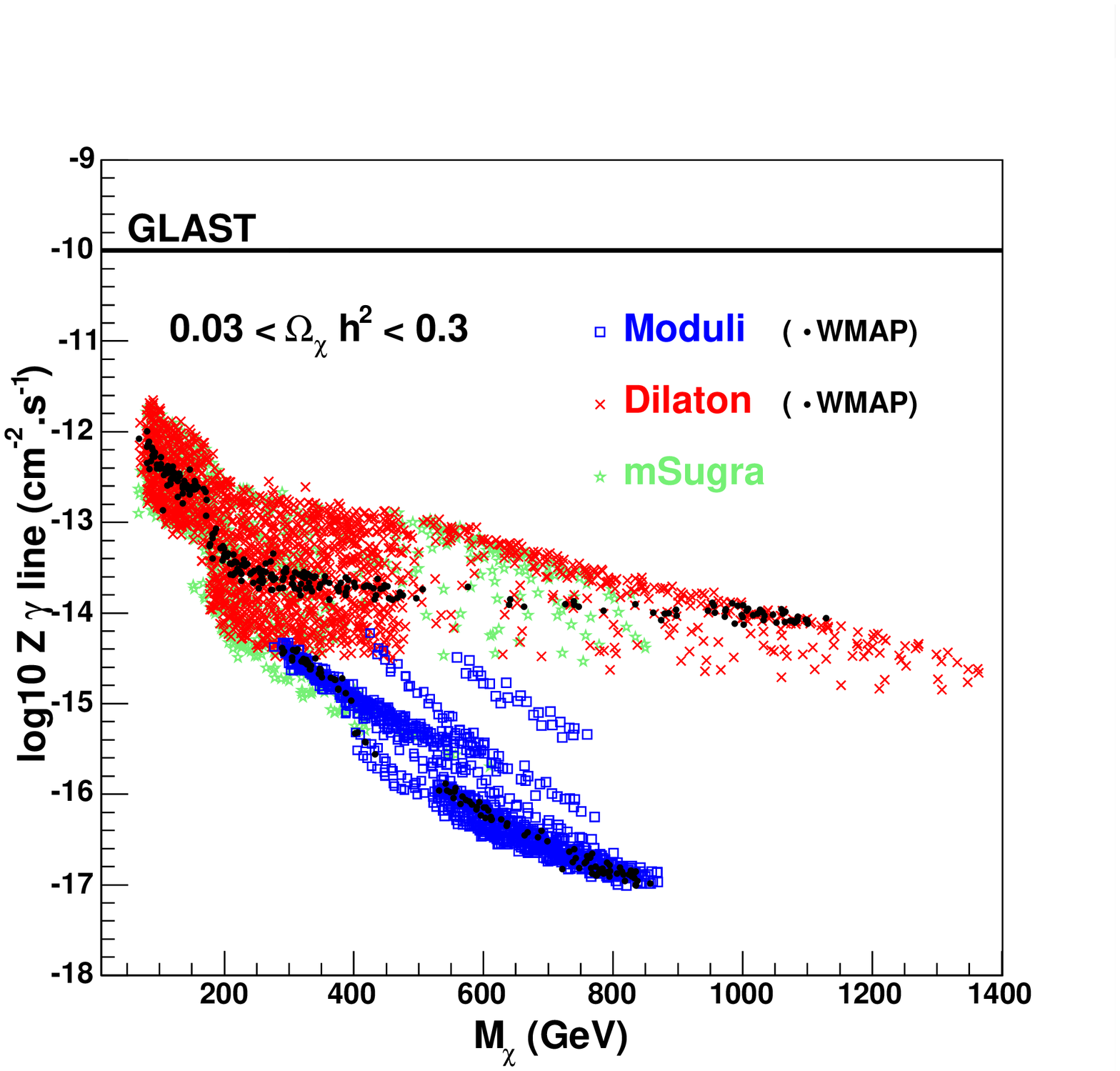}\\
a) & b)
\end{tabular}
\caption{{\footnotesize {\bf The $\gamma \gamma$ and $z \gamma$ lines} versus 
neutralino mass
for tan$\beta=5, 20$ and $35$ taking a NFW profile for the halo,
after a scan in the dilaton parameter space and moduli one 
for $t=0.25,~p=0$, and a comparison with mSugra. The points that violate the 
accelerator constraints are not represented here (see the appendix for the cut applied).}}
\label{fig:gagagazlines}
\end{center}
\end{figure}

In the dilaton dominated breaking scenario, most of the points are excluded
 at low $m_{\chi}$ because they do not respect the accelerator constraints on 
$m_{\chi^+_1}$ (and $m_h$),
whereas heavy neutralinos are forbidden by the closed parameter space 
limited by the gaugino condensation group $\beta$--function 
coefficient $b_+$  
($b_+ \leq b_{E_8}\sim0.57$). This parameter space is larger for lower values
of tan$\beta$ because barely constrained by the realization of the
 electroweak symmetry breaking condition ($\mu^2>0$).
Most of the points with low integrated flux are excluded when
imposing a higher limit on the relic density ($\Omega_{\chi} < 0.3$).
This is due to the correlation between the fluxes and the relic density.
%through the processes of neutralino annihilation. 
In any case,
if we take a lower upper omega limit (like the WMAP one) we keep the points 
with the 
best detection rates. As was pointed out in \cite{Bin2}, the
cosmologically favored region of the parameter space is similar to the
Focus Point (FP) region present in mSUGRA \cite{Feng}, i.e. a mixed higgsino region
leading to strong couplings and high gamma--ray fluxes dominated by $t
\overline{t}$ ($Z$ exchange) or $WW$ ($ZZ$) final state processes. This 
explains the correlation
 between the dilaton and mSUGRA clouds on Fig. \ref{fig:fluxdil}.

In the moduli dominated scenario, the dependence on $\tan \beta$ is not only
due to the mass spectrum or a larger parameter space, but to the annihilation 
processes governing the gamma--ray fluxes. Indeed, 
in the moduli breaking scenario, one can have a relatively 
light pseudo scalar $A$.
The annihilation of the neutralino is thus dominated by the s channel 
$\chi\chi \stackrel{A}{\rightarrow} b\overline{b}$ with a branching ratio approaching
 the 90 percents in most cases (the coupling $Ab\overline{b} 
\propto$ tan$\beta$, 
gives a factor 50 in $\sigma$ between tan$\beta=$ 5 and 35). We can notice
that thanks to the $\delta_{GS}$ parameter the A--pole can be obtained for 
lower value of $\tan{\beta}$ than in mSUGRA.

To study the observability of these fluxes, we have 
estimated the total gamma--ray flux above 60 GeV, and
compared it with the predicted sensitivity of the HESS experiment 
(Figs. \ref{fig:fluxmod}b and \ref{fig:fluxdil}b).
The low neutralino mass region ($ < 400$ GeV)is not accessible to EGRET but will be explored 
by GLAST, whereas HESS will give more precise information for a heavy
neutralino. In that sense, GLAST and HESS experiments are complementary. 
Indeed, evading the NFW
profile assumption and taking a less cuspy density, the GLAST accessible
region of our models is much less overlapping the HESS one. 
For instance, assuming a NFW profile, a non discovery of any signal in HESS 
$and$ GLAST will exclude any dilaton dominated scenario and the high tan$\beta$ 
regime of the moduli domination.

%Furthermore, these experiments could help with the determination of the SUSY 
%parameter tan$\beta$. Actually in the case
%of the moduli scenarios, the relation is almost linear between the gamma flux
%and the $\tan{\beta}$ value through the $\chi \chi \stackrel{A}{\rightarrow}b\bar{b}$
%annihilation channel. This means that a discovery 
%of the LSP with its mass (by another way) cross checked with the 
%value of the gamma flux could give  the value of the tan$\beta$
%parameter. In the case of dilaton domination, any mass of the LSP superior 
%to 600 GeV will be an indication of tan$\beta$ $< 5$.

%For instance, for given $\tan{\beta}$, $m_{\chi}$ (discovered in collider
%experiments) and a given halo, gamma flux results coulds suggest the SUSY 
%breaking mechanism.

%Concerning the comparison with mSugra : some points satisfying experimental
%constraints and within reach of gamma indirect detection experiments with  
%similar phenomenology of the two SUSY breaking scenarios are present : focus 
%point (dilaton like) and stau coannihilation (moduli like) branches.) But the 
%CMSSM/mSugra framework do not give information on parameters driving SUSY 
%breaking. 

Finally, in the framework of effective heterotic models proposed in this
 paper, concerning direct and solar neutrino indirect detection, points
satisfying all constraints in the dilaton case were all accessible to experiments
whereas moduli was not \cite{Bin2}. Concerning gamma indirect detection, the 
conclusion 
depends strongly on the halo profile assumption. For a NFW shape, the dilaton
model is still the most attractive for detection but the moduli one can also be
tested by experiment. In that sense, gamma indirect detection is less an
explicit test of SUSY breaking scenario but a very complementary way to give
 some hints on particles physics hypothesis and dark halo astrophysical 
assumptions.

We close this section by showing $\gamma \gamma$ and $Z \gamma$ line results
on Figs. \ref{fig:gagagazlines}. 
Because dark mater in the halo is extremely non--relativistic,
photons from these processes have an energy width of only 
$\Delta E_{\gamma} / E_{\gamma} \sim 10^{-3}$ and are effectively 
mono--energetic. While this signal would be the most spectacular of
all possible indirect signals, as we can see in Figs. \ref{fig:gagagazlines} the processes being loop-suppressed are much too low to be 
detected, even in any string scenario studied here. 
On another hand, the integrated flux explored before is 
 more observable, but far less 
distinctive and will certainly require additional confirmation to
unambiguously distinguish it from the background or other exotic sources.

\section{Synchrotron radiation}

Another interesting source for the indirect search of DM in the Galactic halo
is the detection of the neutralino via the synchrotron radiation created 
by the propagation of the $e^{\pm}$ products of neutralino annihilation, around the Galactic magnetic field. In our study, we have considered
a magnetic field at equipartition at the Galactic center and constant
elsewhere~\cite{Bertone:2002ms} 
\begin{equation}
B(r)=\mathrm{max}
\[
324\mu\mathrm{G}
\left(
\frac{r}{\mathrm{pc}}
\right)^{-5/4}
,6\mu\mathrm{G}
\]
\end{equation}

\noindent
meaning that the field is constant for galacto-centric distances 
$r > 0.23$ pc.

Lower values of the magnetic field imply a shift of the synchrotron spectrum
toward lower energies. As a consequence, the flux at low frequency will 
increase, favoring the detectability of our models. Nevertheless, we prefer 
to be conservative and consider a $B$ field at equipartition. 

The synchrotron flux per solid angle at a given frequency $\nu$ is given
by~\cite{Bertone:2002ms}

\begin{equation}
L_{\nu}(\psi) \sim \frac{1}{4 \pi}\frac{9}{8}
\left(
\frac{1}{0.29 \pi} \frac{m_e^3 c^4}{e}
\right)^{1/2}
\frac{\sigma v}{M^2}Y_e(M,\nu)\nu^{1/2}I(\psi),
\end{equation}

\noindent
where

\begin{equation}
I(\psi)=\int_{line~of~sight} ds \rho^2(r(s,\psi))B^{-1/2}(r(s,\psi)),
\end{equation}

\noindent
and $s$ is the coordinate running along the line of sight.
$Y_e(M,\nu)$ is the average number of secondary electrons above the energy
$E_m(\nu)$ of the electron giving the maximum contribution at a given 
frequency $\nu$ and given magnetic field $B$. We recall that 
\begin{equation}
E_m(\nu)=
\left(
\frac{4 \pi}{3}\frac{m_e^3 c^4}{e}\frac{\nu}{B}
\right)^{1/2}
\end{equation}

In order to compare our predictions with the observational data, we 
to integrate over the corresponding solid angle, studying two distinct
astrophysical situations and integrating over the string parameters of
the theory in every case :

\begin{itemize}
 
\item{Flux at $\nu=$ 408 MHz in a cone of half width 4 arc sec pointing
around the Galactic Center in a $NFW$ halo model. The observed flux is
$\sim 0.05$ Jy \cite{Davies}.}

\item{Flux at $\nu=$ 327 MHz in a cone of half width 13.5 arc sec pointing
around the Galactic Center in a $NFW$ halo model. The observed flux is
$\sim 362$ Jy \cite{Rosa}.}

\end{itemize}

In our study, we have neglected two processes: self--absorption and absorption 
of electrons by the interstellar medium. In fact, it has been shown in
Ref.~\cite{Bertone:2002ms} that in a NFW profile, the optical depth
$\tau$ can be safely neglected unless very low frequencies are considered 
(of order $\sim 1$ Hz), which is not our case.
Concerning the electron absorption, the authors of~\cite{Bertone:2002ms} observed 
that the absorption coefficient per length is such that
$\alpha_{\nu} < 6~10^{-16} pc^{-1}(B/\mu G)(\nu/GHz)^{-2}$,
justifying our approximation.

%======== FIGURE 9 : Synchrotron Radiation in Dilaton Case for tan beta=35
%after cut======

\begin{figure}
%\begin{center}
\begin{tabular}{cc}
 \includegraphics[width=0.45\textwidth]{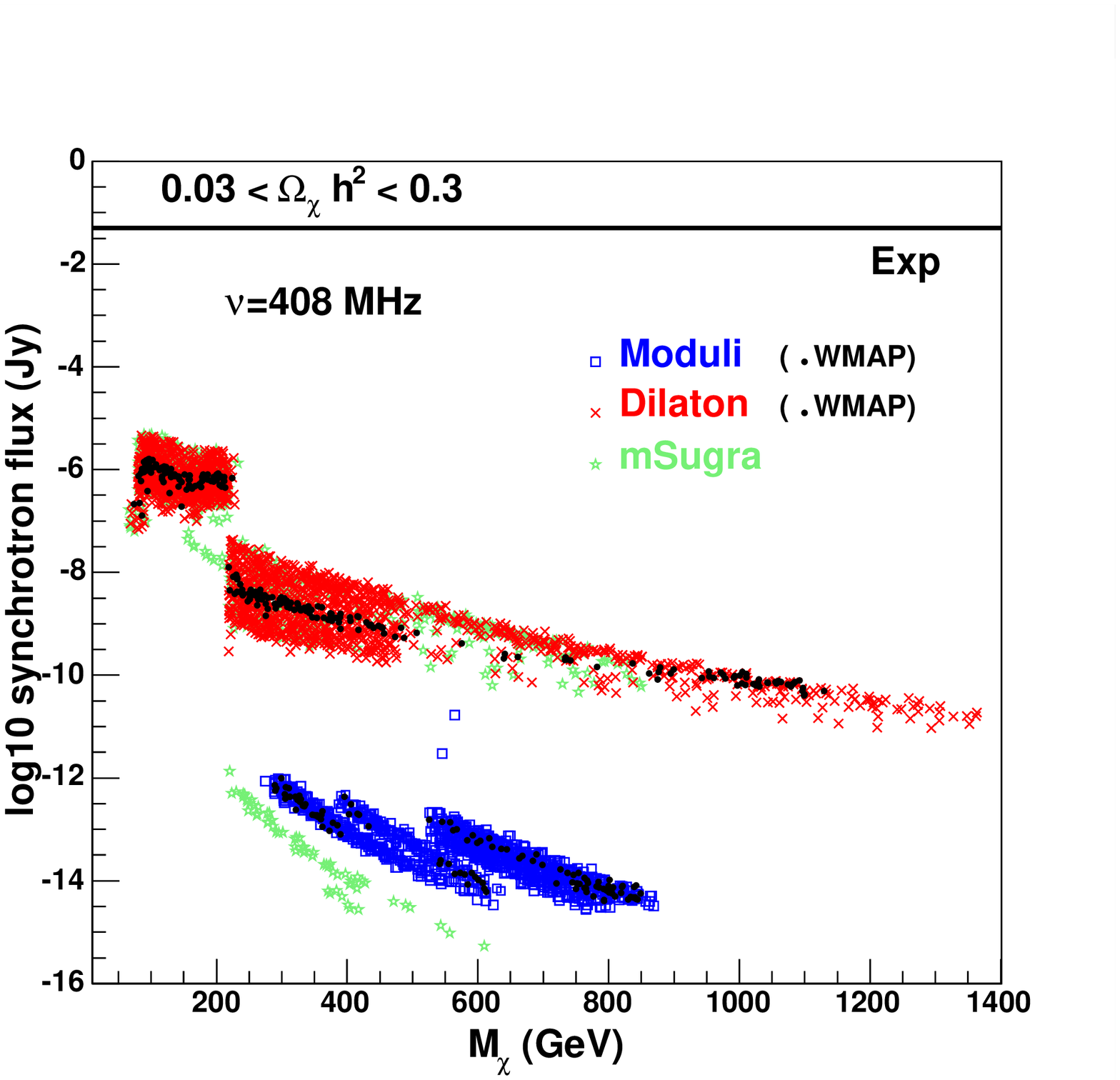}&
\includegraphics[width=0.45\textwidth]{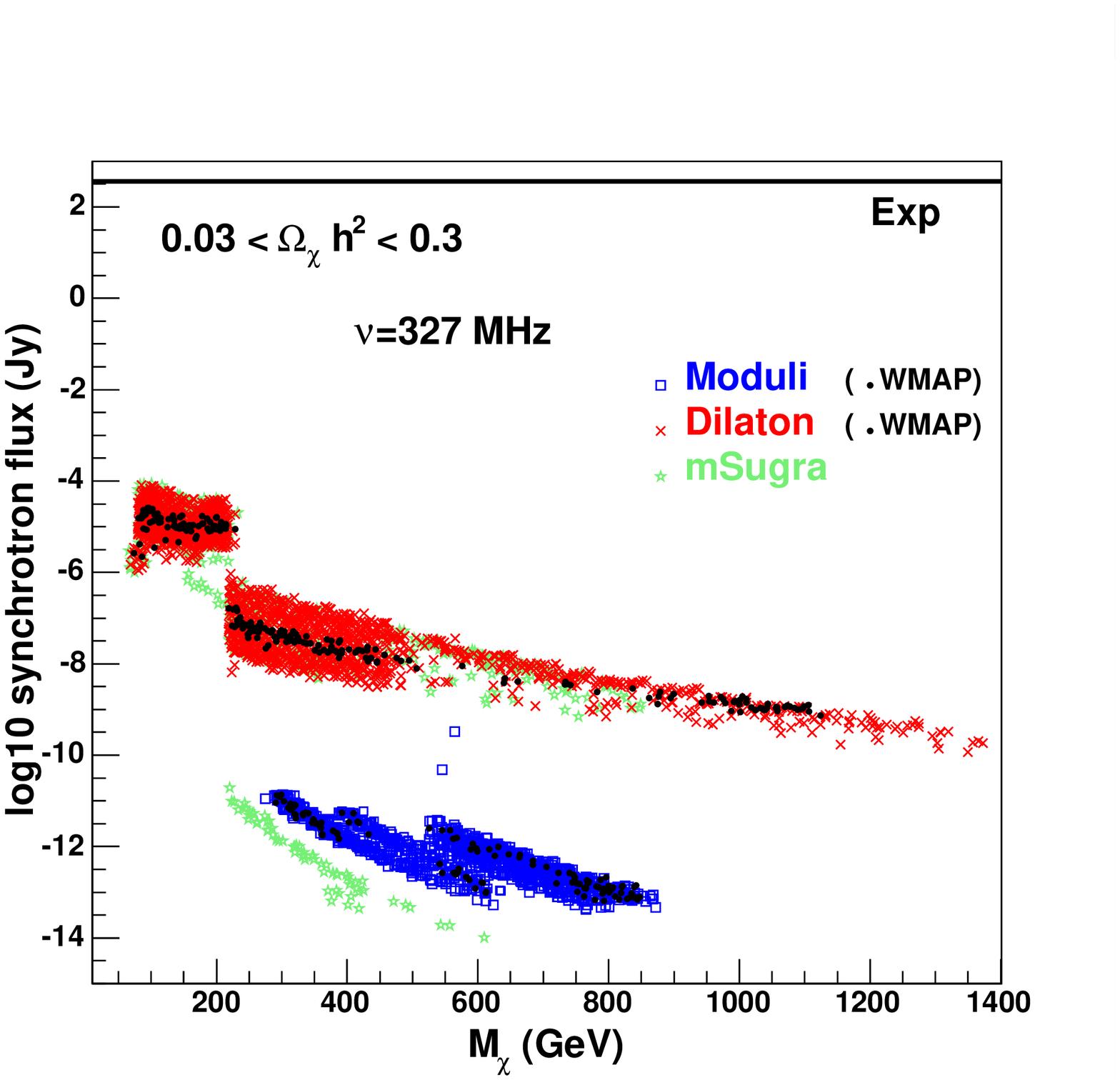}\\
a) & b)
\end{tabular}
\caption{{\footnotesize {\bf Predicted synchrotron radiation from
the Galactic center} versus the neutralino mass for a NFW profile in a moduli dominated
scenario and a dilaton one for tan$\beta=5,20,35$ (after experimental cuts)
at a frequency of 408 MHz (left) and 327 MHz (right) compared to the flux 
observed limit at the same frequencies (horizontal lines).  We also show the results of a ``corresponding'' mSugra scan}}
\label{fig:Synchrotron}
%\end{center}
\end{figure}

%==============================================================================

In any case, for a NFW halo profile, the fluxes obtained for the models
considered here do not constrain the mass of the DM particle, contrary to 
what happens in the case of Kaluza-Klein particles~\cite{Bertone:2002ms}. 
Naturally, the synchrotron flux shapes we
obtain are similar to what we obtain for $\gamma$ fluxes due to the same
$\sigma v$ dependence, but the resulting values in the case of a NFW 
profile are several orders of magnitudes lower than experimental constraints.
Recently, the calculation of synchrotron radiation from neutralino 
annihilation has been revisited ~\cite{Aloisio:2004hy}. Unfortunately
the processes described there are not sufficient to make the flux in
our models observable.  

\section{Conclusions}

%$\rightarrow$ The parameter space in both model is larger than in mSUGRA case after 
%that the experimental constraints are applied

We have studied gamma-ray and synchrotron emission from the Galactic center, 
in the context of an effective string inspired framework, and discussed two
scenarios: in the first one, the SUSY breaking is transmitted by the compactification
moduli $T^{\alpha}$; while in second one, it is transmitted by the dilaton field
$S$, via their respective auxiliary fields. 
Typically, models in the dilaton dominated SUSY breaking scenario 
lead to a higher annihilation rate than the moduli scenario. Concerning the
continuum gamma-ray flux, both scenarios
are within the reach of the experimental sensitivities of GLAST and HESS for a NFW
halo profile. For the same profile, the gamma-ray line signal is suppressed and
beyond the experimental sensitivity. 

The synchrotron emission is too low to be constrained 
by experiments even with a more cuspy profile. Due to the dependence  of the
prospects of detection on both theoretical high energy physics assumptions and astrophysical
parameters of the halo dark matter distribution, gamma-ray indirect detection of
neutralino dark matter can give interesting information either on
astrophysical hypotheses or on the SUSY breaking scenario by the complementarity
with other experimental searches like direct detection or neutrino telescopes
(studied in \cite{Bin2} in the same context) or accelerators.

%$\rightarrow$ Moduli in general more difficult for the observation than dilaton

%$\rightarrow$ Under astrophysical assumption and by complementarity with other
%experiment searches neutralino dark matter gamma indirect detection can be a
%tools to distinguish the two kind of breaking mechanism

%$\rightarrow$ A word on synchrotron and monochromatic rays : out of reach in NFW. 

\vspace{1cm}

{\bf {\large Acknowledgments}}

E.N. would like to thank J. Orloff for useful discussions. G.B. is supported by 
the DOE and the NASA grant NAG 5-10842 at Fermilab. Y.M. thanks C. Mu\~noz
for valuable discussions and A.M. Teixeira for the help provided during 
this work. Y.M was supported by the European Union under contract HPRN-CT-2000-00148.

%\vspace{1cm}

\newpage

\appendix
\noindent {\bf }
\label{appendixconstrain}
\vspace{0.5cm}

\section{Annihilation branching ratios for $\tan{\beta=5}$.}

\begin{figure}[h]
\begin{center}
 \includegraphics[width=0.8\textwidth]{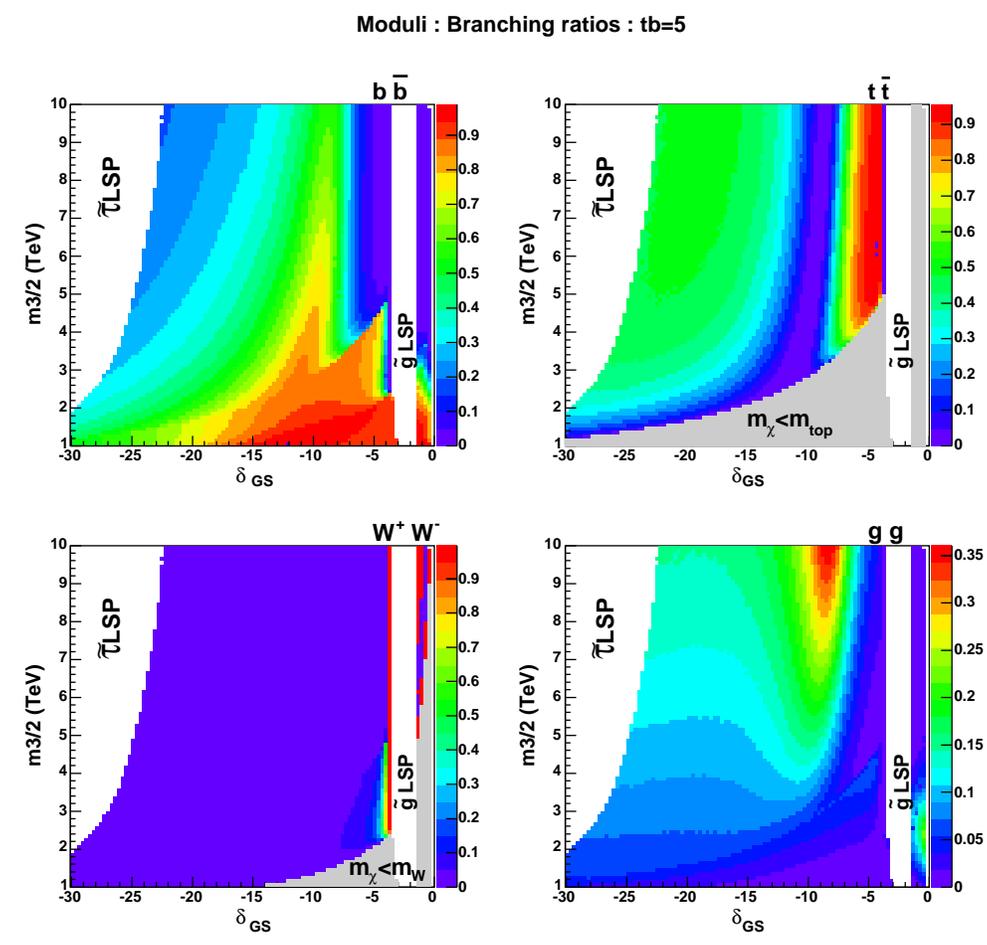}
\caption{\small Moduli domination regime: dominant annihilation branching ratios in the
  ($\delta_{GS},m_{3/2}$) plane for $\tan{\beta}=5,~t=0.25,~p=0$. Regions with gluino and
  stau LSP are indicated. We also show in grey the kinematically forbidden region for
  each channel.}
\label{fig:BR-modtb5}
\end{center}
\end{figure}

\begin{figure}[h]
\begin{center}
 \includegraphics[width=0.8\textwidth]{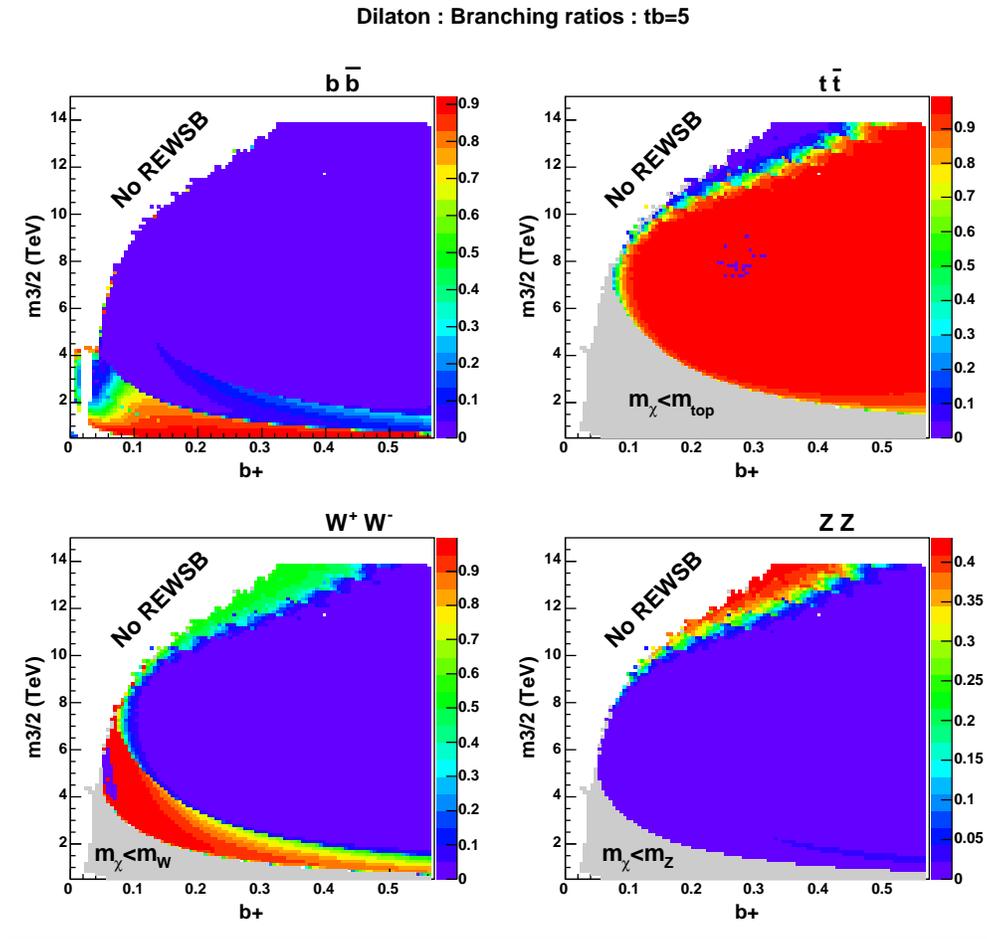}
\caption{\small  Dominant annihilation branching ratios in the
  ($b_{+},m_{3/2}$) plane for $\tan{\beta}=5,~t=0.25,~p=0$. Regions where radiative
  electroweak symmetry breaking can not occur are indicated (No REWSB). We
  also show in grey the kinematically forbidden region for
  each channel.}
\label{fig:BR-diltb5}
\end{center}
\end{figure}

We show here the figures corresponding to Fig.\ref{fig:BR-modtb35} and
\ref{fig:BR-diltb35} for  $\tan{\beta=5}$. The case of moduli domination regime
is illustrated in Fig.\ref{fig:BR-modtb5}, while in Fig.~\ref{fig:BR-diltb5} we
show the case of dilaton dominated SUSY breaking.

%\subsection{Moduli dominated SUSY breaking}

%==============================================================================

%==============================================================================

%\newpage

%\subsection{Dilaton dominated SUSY breaking}

%==============================================================================

%==============================================================================

\newpage

\section{Parameter ranges}

We performed a scan in the moduli and dilaton parameter space with the 
following values :

\begin{itemize}

\item Moduli : 

$t=0.25,p=0$

$\tan{\beta}=5$ ; $ 0<m_{3/2}< 10000 $ GeV ; $-30<\delta_{\mathrm GS}<0$

$\tan{\beta}=20$ ; $ 0<m_{3/2}< 16000 $ GeV ; $-20<\delta_{\mathrm GS}<0$

$\tan{\beta}=35$ ; $ 0<m_{3/2}< 20000 $ GeV ; $-10<\delta_{\mathrm GS}<0$

\item Dilaton: 

$t=0.25,p=0$

$\tan{\beta}=5$ ; $ 0<m_{3/2}< 14000 $ GeV ; $0<b_+<0.57$

$\tan{\beta}=20$ ; $ 0<m_{3/2}< 4500 $ GeV ; $0<b_+<0.57$

$\tan{\beta}=35$ ; $ 0<m_{3/2}< 4000 $ GeV ; $0<b_+<0.57$,

\end{itemize}

\noindent
corresponding to our previous study on direct detection and neutrino indirect
detection \cite{Bin2}. We also show for comparison an equivalent mSugra cloud
with the following parameter :

\begin {itemize}

\item mSugra : 

$A_0=0$

$\tan{\beta}=5$ ; $ 0<m_0< 14000 $ GeV ; $ 0<m_{1/2}< 2000 $ GeV

$\tan{\beta}=20$ ; $ 0<m_0< 4000 $ GeV ; $ 0<m_{1/2}< 1500 $ GeV

$\tan{\beta}=35$ ; $ 0<m_0< 3000 $ GeV ; $ 0<m_{1/2}< 1000 $ GeV

\end{itemize}

\begin{figure}
\begin{center}
\begin{tabular}{cc}
 \includegraphics[width=0.45\textwidth]{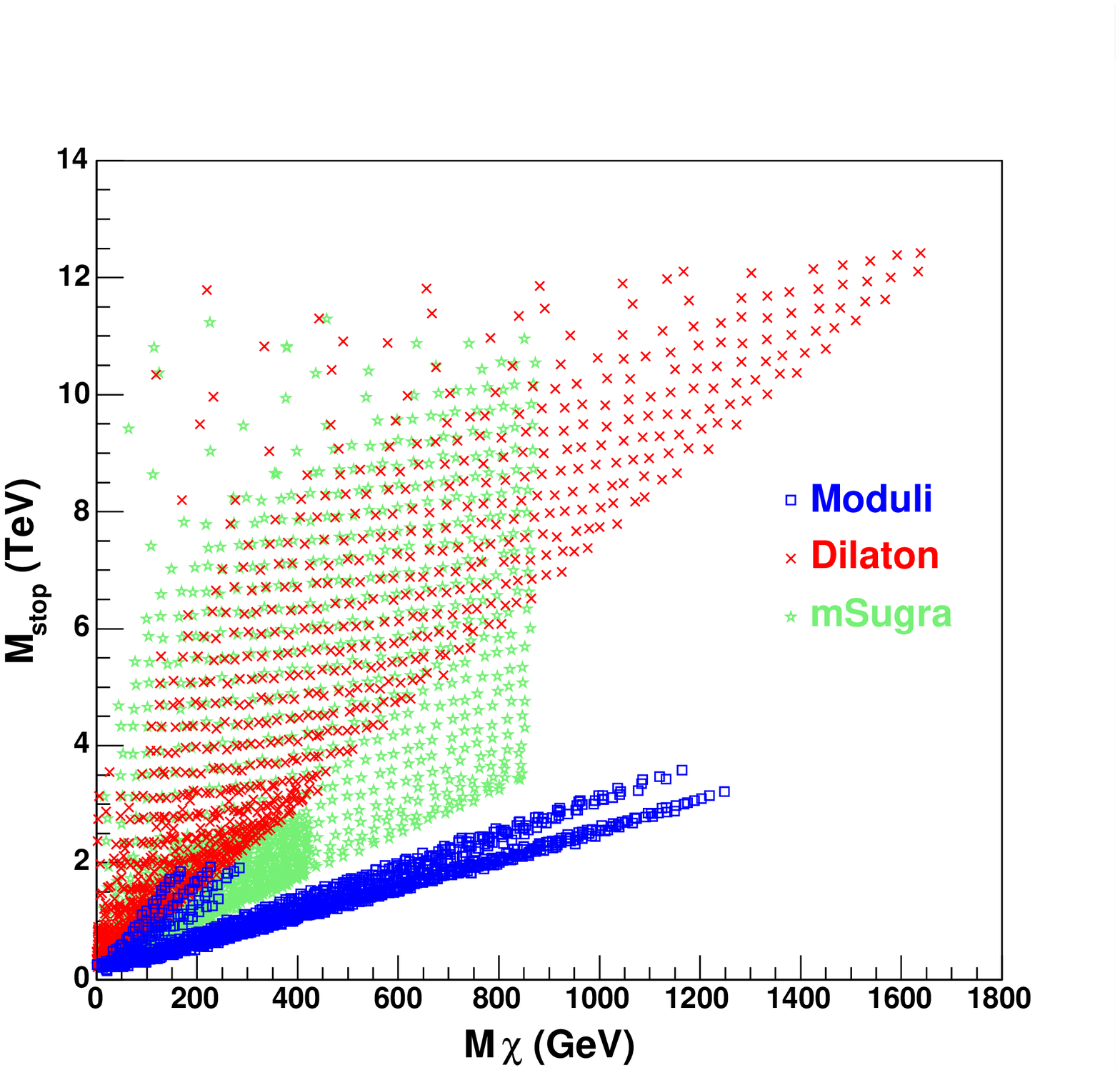}&
\includegraphics[width=0.45\textwidth]{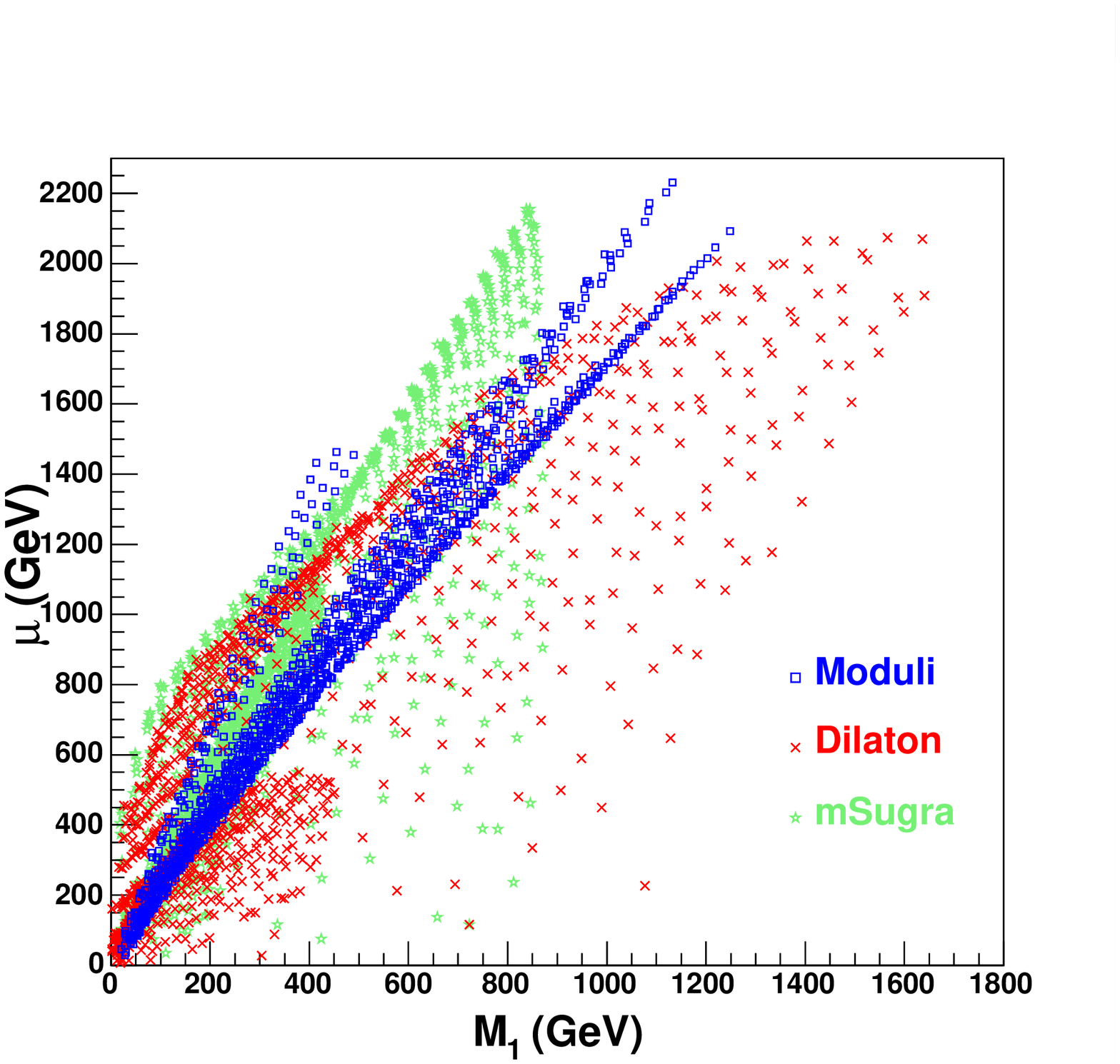}\\
a) & b)
\end{tabular}
\caption{{\footnotesize Low energy $(m_{\chi},m_{\tilde{t}})$ and $(M_1,\mu)$
    planes showing that the parameter space we choose are quite equivalent.}}
\label{fig:lownrjplanes}
\end{center}
\end{figure}

We show on Fig. \ref{fig:lownrjplanes} two typical low energy 
planes:  $(m_{\chi},m_{\tilde{t}})$ and $(M_1,\mu)$ for $\tan{\beta}=5,35$ only. This illustrates that our
choices of parameter ranges are coherent from the low energy point of view and
allow for comparison between the different models. An extended study of such
comparison is the subject of a forthcoming paper \cite{Bin4}.

\section{Experimental constraints}

We apply the following conservative cuts on our models:

\begin{itemize}
\item Higgs mass: $ m_h > 113.5$ GeV \cite{Higgslimit},

\item Chargino mass: $m_{\chi^+} > 103.5$  GeV \cite{charginolimit},

\item Relic density: $0.03<\Omega_{\chi}{\rm h}^2<0.3$, 
but we also show the WMAP~\cite{WMAP} range $
\Omega^{WMAP}_{\rm CDM} {\rm h}^2 = 0.1126^{+0.0161}_{-0.0181}$,

\item $ b \rightarrow s \gamma$ Constraint  \cite{bench}:\\ 
$2.33 \times 10^{-4} < \mathrm{BR} (b \rightarrow s \gamma) < 4.15 
\times 10^{-4}$,

\item The muon anomalous magnetic moment \cite{Davier:2003pw}: $ -11.6~<~\delta_{\mu}^{\mathrm{new \, physics}} =  \delta_{\mu}^{\mathrm{exp}}-\delta_{\mu}^{\mathrm{SM}}~<~30.4 ~~~~~
 [2~\sigma] $.

\end{itemize}

\section{Tools}

For our computations we have interfaced SUSPECT\cite{Suspect}, MICROMEGAS\cite{Micromegas} and
DARKSUSY\cite{Darksusy}.

Concerning MSSM renormalization group equations and radiative electroweak
symmetry breaking we use the code SUSPECT. It includes 2-loops RGE evolution
from high scale down to low energy scale (we took
$Q=\sqrt{\tilde{t}_1\tilde{t}_2}$ \cite{GaRiZw90}) and minimizes the 1-loop scalar
potential by solving iteratively the condition

\begin{equation}
{\mu}^{2}=\frac{\(m_{H_d}^{2}+\delta m_{H_d}^{2}\) - 
  \(m_{H_u}^{2}+\delta m_{H_u}^{2}\) \tan{\beta}}{\tan^{2}{\beta}-1} 
-\frac{1}{2} M_{Z}^{2} ,
\end{equation}

\noindent
giving the $\mu$ parameter which drives the crucial neutralino higgsino fraction for dark matter studies.

We then transfer all MSSM parameters to the code MICROMEGAS which calculates the
full physical mass spectrum including radiative corrections on Higgs masses
and widths. The calculation of neutralino relic density is then achieved
 following the iterative procedure described in~\cite{Gondolo:1990dk}
and
 including all annihilation and co-annihilation processes by solving 

\begin{equation}
\frac{dY}{dT}=\sqrt{\pi g_*(T)/45G}\langle \sigma v \rangle (Y^2-Y^2_eq)
\end{equation}
where $Y$ is the abundance at the temperature $T$, $g_*$ a degrees of freedom
coming from thermodynamics, $G$ the Newton constant and $\langle \sigma v
\rangle$ the thermally average cross section of all processes concerning $Y$.

Masses and couplings are then entered in DARKSUSY which calculates the
indirect detection rates of species $i=\gamma, e^+ ...$ coming from the
Galactic Centre for a chosen galactic halo profile by splitting the particle
physics dependance ($\propto \langle \sigma v \rangle d\phi/dE_i$, $\langle
\sigma v \rangle$ including
here only annihilations) and the
astrophysics part {\it i.e} the integration along the line of sight ($\int ds
\rho^2(r(s,\psi))$, $\rho(r)$ being the dark matter density distribution). 
%in a $10^-3$ degree cone around the galactic centre direction ($\psi=0$).

\nocite{}
%\begin{bibliography}
%\bibliographystyle{unsrt}


\begin{thebibliography}{99}

%\cite{Bertone:2004pz}
\bibitem{Bertone:2004pz}
G.~Bertone, D.~Hooper and J.~Silk,
%``Particle Dark Matter: Evidence, Candidates and Constraints,''
arXiv:hep-ph/0404175.
%%CITATION = HEP-PH 0404175;%%

\bibitem{mSUGRA}
  {\rm J.~R.~Ellis, T.~Falk, G.~Ganis, K.~A.~Olive and M.~Srednicki},
  {\it Phys. Lett.} {\bf B510} {\rm (2001) 236}; \\
  {\rm J.~R.~Ellis, K.~A.~Olive and Y.~Santoso},
  {\it New Jour. Phys.} {\bf 4} {\rm (2002) 32}; \\
  {\rm L.~Roszkowski, R.~Ruiz de Austri and T.~Nihei},
  {\it JHEP} {\bf 0108} {\rm (2001) 024}; \\
  {\rm A.~Djouadi, M.~Drees and J.~L.~Kneur},
  {\it JHEP} {\bf 0108} {\rm (2001) 055}; \\
  {\rm H.~Baer, C.~Balazs and A.~Belyaev},
  {\it JHEP} {\bf 0203} {\rm (2002) 042}.\\
  {\rm U.~Chattopadhyay, A.~Corsetti and P.~Nath},
  {\it Phys.\ Rev.}\ D {\bf 68}, 035005 (2003).\\
  {\rm H.~Baer, A.~Belyaev, T.~Krupovnickas and J.~O'Farrill},
  arXiv:hep-ph/0405210.


 \bibitem{EllisHiggs}
 {\rm J.~R.~Ellis, A.~Ferstl, K.~A.~Olive and Y.~Santoso},
 {\it Phys.\ Rev.}\ D {\bf 67}, 123502 (2003)
 
\bibitem{Bottino1}
  {\rm  V.~Berezinsky, A.~Bottino, J.~Ellis, N.~Fornengo, G.~Mignola,
S.~Scopel}, 
  {\it Astropart.Phys.} {\bf 5} {\rm (1996) 1-26},
{arXiv:hep-ph/9508249}.

\bibitem{Arnowitt1}
  {\rm   P.~Nath, R.~Arnowitt}, 
  {\it Phys.Rev.} {\bf D56} {\rm (1997) 2820-2832},
{arXiv:hep-ph/9701301}. 

\bibitem{Mynonuniv}
  {\rm   V.~Bertin, E.~Nezri, J.~Orloff}, 
  {\it JHEP} {\bf 02} {\rm (2003) 046},
{arXiv:hep-ph/0210034}.

\bibitem{BirkedalnonU}
{\rm A.~Birkedal-Hansen B.~D.~Nelson}
{\it Phys.Rev.} {\bf D67} {\rm (2003) 095006},
{arXiv:hep-ph/0211071}.

\bibitem{Nath2}
  {\rm   A.~Corsetti, P.~Nath}, 
  {\it Phys.Rev} {\bf D64} {\rm (2001) 125010},
{arXiv:hep-ph/0003186}.
%\cite{Profumo:2003em}
\bibitem{Profumo:2003em}
S.~Profumo,
 %``Neutralino dark matter, b - tau Yukawa unification and non-universal
%sfermion masses,''
Phys.\ Rev.\ D {\bf 68}, 015006 (2003)
%[arXiv:hep-ph/0304071].
%%CITATION = HEP-PH 0304071;%%

%\cite{Ullio:qe}
\bibitem{Ullio:qe}
P.~Ullio,
 %``Indirect Searches For Neutralino dark matter Candidates In Anomaly-Mediated
%Supersymmetry Breaking Scenarios,''
Nucl.\ Phys.\ Proc.\ Suppl.\  {\bf 110}, 82 (2002).
%%CITATION = NUPHZ,110,82;%%


%\cite{Cesarini:2003nr}
\bibitem{Cesarini:2003nr}
A.~Cesarini, F.~Fucito, A.~Lionetto, A.~Morselli and P.~Ullio,
%``The galactic center as a dark matter gamma-ray source,''
arXiv:astro-ph/0305075.
%%CITATION = ASTRO-PH 0305075;%%


%\cite{Hooper:2003ka}
\bibitem{Hooper:2003ka}
D.~Hooper and L.~T.~Wang,
 %``Direct and indirect detection of neutralino dark matter in selected
%supersymmetry breaking scenarios,''
Phys.\ Rev.\ D {\bf 69}, 035001 (2004)
[arXiv:hep-ph/0309036].
%%CITATION = HEP-PH 0309036;%%


%\cite{Bottino:2004qi}
\bibitem{Bottino:2004qi}
A.~Bottino, F.~Donato, N.~Fornengo and S.~Scopel,
 %``Indirect signals from light neutralinos in supersymmetric models without
%gaugino mass unification,''
arXiv:hep-ph/0401186.
%%CITATION = HEP-PH 0401186;%%

%\cite{Cerdeno:2004zj}
\bibitem{Cerdeno:2004zj}
D.~G.~Cerdeno and C.~Munoz,
%``Neutralino dark matter in supergravity theories with non-universal scalar and
%gaugino masses,''
arXiv:hep-ph/0405057.
%%CITATION = HEP-PH 0405057;%%

%\cite{Chattopadhyay:2004dt}
\bibitem{Chattopadhyay:2004dt}
U.~Chattopadhyay and P.~Nath,
%``Modular invariant soft breaking, WMAP, dark matter and sparticle mass
%limits,''
arXiv:hep-ph/0405157.
%%CITATION = HEP-PH 0405157;%%

%\cite{Profumo:2004ty}
\bibitem{Profumo:2004ty}
S.~Profumo and P.~Ullio,
%``The Role of Antimatter Searches in the Hunt for Supersymmetric Dark Matter,''
arXiv:hep-ph/0406018.
%%CITATION = HEP-PH 0406018;%%


\bibitem{BiGaNe01} 
  {\rm P.~Bin\'etruy, M.~K.~Gaillard and B.~D.~Nelson}, 
  {\it Nucl. Phys.} {\bf B604} {\rm (2001) 32}. 

\bibitem{Bin1}
{\rm P.~Bin\'etruy, A.~Birkedal-Hansen, Y.~Mambrini, B.D.~Nelson}
{arXiv:hep-ph/0308047}.

\bibitem{Bin2}
{\rm P.~Bin\'etruy, Y.~Mambrini, E.~Nezri}
{arXiv:hep-ph/0312155}.

\bibitem{NelsonTeva}
{\rm G.L.~Kane, J.~Lykken, S.~Mrenna, B.D.~Nelson, L.T.~Wang, T.T.~Wang}
{\it Phys.Rev.} {\bf D67} {\rm (2003) 045008} 
{arXiv:hep-ph/0209061}.


\bibitem{BiGaWu96} 
  {\rm P.~Bin\'{e}truy, M.~K.~Gaillard and Y.-Y.~Wu}, 
  {\it Nucl. Phys.} {\bf B481} {\rm (1996) 109}. 

\bibitem{BiGaWu97a} 
  {\rm P.~Bin\'{e}truy, M.~K.~Gaillard and Y.-Y.~Wu}, 
  {\it Nucl. Phys.} {\bf B493} {\rm (1997) 27}. 

\bibitem{GaNeWu99} 
  {\rm M.~K.~Gaillard, B.~D.~Nelson and Y.-Y.~Wu}, 
  {\it Phys. Lett.} {\bf B459} {\rm (1999) 549}. 

\bibitem{GaNe00b} 
  {\rm M.~K.~Gaillard and B.~D.~Nelson}, 
  {\it Nucl. Phys.} {\bf B588} {\rm (2000) 197}. 
\bibitem{LEPWG}{\rm ALEPH collaboration}{\it Phys.Lett.} {\bf B583} {\rm (2004) 247-263}


%\cite{Falvard:2002ny}
%\bibitem{Falvard:2002ny}
%A.~Falvard {\it et al.},
 %``Supersymmetric dark matter in M31: Can one see neutralino annihilation with
%CELESTE?,''
%Astropart.\ Phys.\  {\bf 20}, 467 (2004)
%[arXiv:astro-ph/0210184].
%%CITATION = ASTRO-PH 0210184;%%

 \bibitem{WMAP}
  C.~L.~Bennett {\it et al.}, 
{\it Astrophys.J.Suppl.} {\rm 148:1,2003}\\
%arXiv:astro-ph/0302207;\\
  D.~N.~Spergel {\it et al.}, 
{\it Astrophys.J.Suppl.} {\rm 148:175,2003}
%arXiv:astro-ph/0302209.


%\cite{Navarro:1996he}
\bibitem{Navarro:1996he}
J.~F.~Navarro, C.~S.~Frenk and S.~D.~M.~White,
%``A Universal density profile from hierarchical clustering,''
Astrophys.\ J.\  {\bf 490} (1997) 493.
%%CITATION = ASJOA,490,493;%%

\bibitem{Kravtsov:1997dp}
A.~V.~Kravtsov, A.~A.~Klypin, J.~S.~Bullock and J.~R.~Primack,
%``The Cores of dark matter Dominated Galaxies: theory vs. observations,''
arXiv:astro-ph/9708176.
%%CITATION = ASTRO-PH 9708176;%%

%\cite{Moore:1999gc}
\bibitem{Moore:1999gc}
B.~Moore, T.~Quinn, F.~Governato, J.~Stadel and G.~Lake,
%``Cold collapse and the core catastrophe,''
Mon.\ Not.\ Roy.\ Astron.\ Soc.\  {\bf 310} (1999) 1147
[arXiv:astro-ph/9903164].
%%CITATION = ASTRO-PH 9903164;%%


\bibitem{Bergstrom:1997fj}
L.~Bergstrom, P.~Ullio and J.~H.~Buckley,
 %``Observability of gamma rays from dark matter neutralino annihilations  in
%the Milky Way halo,''
Astropart.\ Phys.\  {\bf 9} (1998) 137
[arXiv:astro-ph/9712318].
%%CITATION = ASTRO-PH 9712318;%%

\bibitem{schodel} R.~Sch\"odel et al. Nature 419, 694 (2002)

\bibitem{Gondolo:1999ef}
P.~Gondolo and J.~Silk,
%``dark matter annihilation at the galactic center,''
Phys.\ Rev.\ Lett.\  {\bf 83} (1999) 1719
[arXiv:astro-ph/9906391].
%%CITATION = ASTRO-PH 9906391;%%

\bibitem{Bertone:2002je}
G.~Bertone, G.~Sigl and J.~Silk,
%``Annihilation Radiation from a dark matter Spike at the Galactic Centre,''
Mon.\ Not.\ Roy.\ Astron.\ Soc.\  {\bf 337} (2002) 98
[arXiv:astro-ph/0203488].
%%CITATION = ASTRO-PH 0203488;%%

\bibitem{deblok} de Blok, W. J. G., McGaugh, S. S., Bosma, A., 
and Rubin, V. C., ApJ 552, L23

%\cite{Swaters:2002rx}
\bibitem{Swaters:2002rx}
R.~A.~Swaters, B.~F.~Madore, F.~C.~V.~Bosch and M.~Balcells,
 %``The Central Mass Distribution in Dwarf and Low Surface Brightness
%Galaxies,''
Astrophys.\ J.\  {\bf 583} (2003) 732
[arXiv:astro-ph/0210152].
%%CITATION = ASTRO-PH 0210152;%%

\bibitem{vandenBosch:1999ka}
F.~C.~van den Bosch, B.~E.~Robertson, J.~J.~Dalcanton and W.~J.~G.~de Blok,
 %``Constraints on the Structure of dark matter Halos from the Rotation Curves
%of Low Surface Brightness Galaxies,''
arXiv:astro-ph/9911372.
%%CITATION = ASTRO-PH 9911372;%%

%\cite{Hayashi:2003sj}
\bibitem{Hayashi:2003sj}
E.~Hayashi {\it et al.},
 %``The Inner Structure of LCDM Halos II: Halo Mass Profiles and LSB Rotation
%Curves,''
arXiv:astro-ph/0310576.
%%CITATION = ASTRO-PH 0310576;%%

%\cite{Bertone:2002ms}
\bibitem{Bertone:2002ms}
G.~Bertone, G.~Servant and G.~Sigl,
%``Indirect detection of Kaluza-Klein dark matter,''
Phys.\ Rev.\ D {\bf 68} (2003) 044008
[arXiv:hep-ph/0211342].
%%CITATION = HEP-PH 0211342;%%

 \bibitem{Jungman} G. Jungman, M. Kamionkowski and K. Griest, 
Phys. Rept. 267 (1996) 195.  

%\cite{Barwick:1997ig}
\bibitem{Barwick:1997ig}
S.~W.~Barwick {\it et al.}  [HEAT Collaboration],
%``Measurements of the cosmic-ray positron fraction from 1-GeV to 50-GeV,''
Astrophys.\ J.\  {\bf 482}, L191 (1997)
[arXiv:astro-ph/9703192].
%%CITATION = ASTRO-PH 9703192;%%

\bibitem{Feng}
  {\rm   J.L.~Feng, K.T.~Matchev, F.~Wilczek},
  {\it Phys.Rev.} {\bf D63} {\rm (2001) 045024},
{arXiv:astro-ph/0008115}.

\bibitem{Bin4}
{\rm P.~Bin\'etruy, Y.~Mambrini, E.~Nezri}
in preparation.


\bibitem{Higgslimit}
  {\rm ALEPH Collaboration (A. Heister et al.)}
  {\it Phys.Lett.} {\bf B526} {\rm (2002) 191}.
\bibitem{charginolimit}
  {\rm ALEPH Collaboration (A. Heister et al.)}
  {\it Phys.Lett.} {\bf B533} {\rm (2002) 223}.
\bibitem{bench}
  {\rm M.~Battaglia, A.~De~Roeck, J.~Ellis, F.~Gianotti, K.~T.~Matchev
  K.~A.~Olive, L.~Pape, G.~Wilson},
  {\it Eur. Phys. J.} {\bf C22} {\rm (2001) 535}.
%\cite{Davier:2003pw}
\bibitem{Davier:2003pw}
M.~Davier, S.~Eidelman, A.~Hocker and Z.~Zhang,
 %``Updated estimate of the muon magnetic moment using revised results from e+
%e- annihilation,''
Eur.\ Phys.\ J.\ C {\bf 31} (2003) 503
[arXiv:hep-ph/0308213].
%%CITATION = HEP-PH 0308213;%%
\bibitem{Suspect}
  {\rm A.~Djouadi, J.~L.~Kneur and G.~Moultaka},
  {\it SuSpect: a Fortran Code for the Supersymmetric and Higgs Particle
  Spectrum in the MSSM},
  {\rm hep-ph/0211331},
  {\tt http://www.lpm.univ-montp2.fr:6714/\char126kneur/suspect.html}.
\bibitem{Micromegas}
  {\rm G.~Belanger, F.~Boudjema, A.~Pukhov and A.~Semenov},
  {\it MicrOMEGAs: A Program for Calculating the Relic Density in the MSSM},
  {\it Comput.Phys.Commun.} {\bf 149} {\rm (2002) 103-120},
  {\rm hep-ph/0112278},\\
  {\tt http://wwwlapp.in2p3.fr/lapth/micromegas}.
\bibitem{Darksusy}
P.~Gondolo, J.~Edsjo, P.~Ullio, L.~Bergstrom, M.~Schelke and E.~A.~Baltz,
{\it DarkSUSY: A numerical package for supersymmetric dark matter calculations},astro-ph/0211238.
P.~Gondolo, J.~Edsjo, L.~Bergstrom, P.~Ullio, et T.~Baltz, {\it DarkSusy
  program, http://www.physto.se/~edsjo/darksusy/}
\bibitem{GaRiZw90}
  {\rm G.~Gamberini, G.~Ridolfi and F.~Zwirner},
  {\it Nucl. Phys.} {\bf B331} {\rm (1990) 331}.
\bibitem{Gondolo:1990dk}
P.~Gondolo and G.~Gelmini,
%``Cosmic Abundances Of Stable Particles: Improved Analysis,''
Nucl.\ Phys.\ B {\bf 360} (1991) 145.
%%CITATION = NUPHA,B360,145;%%

\bibitem{Davies} {\rm R.~D.~ Davies, D.~ Walsh and R.~S.~ Booth}, 
{\it Mon.~ Not.~ Roy.~ Astron.~ Soc.~}, {\bf 177} {\rm (1976) 319}
\bibitem{Rosa} {\rm T.~N.~ La Rosa, N.~E.~ Kassim, T.~J.~ Lazio and S.~D.~ Hyman}, 
{\it Astron.~ Journ.~}, {\bf 119} {\rm (2000) 207}
\bibitem{Ullio3} {\rm L.~ Bergstrom, P.~ Ullio and J.~H.~ Buckley, 
~"Observability of gamma rays from dark matter neutralino annihilations in
the Milky Way halo,"}, 
{\it Astropart.~ Phys.~}, {\bf 9} {\rm (1998) 137 ~[astro--ph/9712318].}

\bibitem{Aloisio:2004hy}
R.~Aloisio, P.~Blasi and A.~V.~Olinto,
%``Neutralino annihilation at the galactic center revisited,''
arXiv:astro-ph/0402588.
%%CITATION = ASTRO-PH 0402588;%%




\end{thebibliography}
\end{document}